\documentclass[10pt]{article}
\usepackage{bbm}
\usepackage[final]{graphics}
\usepackage{amsmath}
\usepackage{amsfonts,amsbsy}
\usepackage{amssymb}
\usepackage{color}
\definecolor{linkcolor}{rgb}{0.4,0.1,0.1}
\definecolor{bibcolor}{rgb}{0.4,0.1,0.1}
\usepackage[colorlinks,linkcolor=linkcolor,citecolor=bibcolor,pdfpagelabels,hyperindex=true]{hyperref}

\def\empile#1\over#2{\mathrel{\mathop{\kern 0pt#1}\limits_{#2}}}
\def\bs{\boldsymbol}

\newcommand{\slv}{\raise.15ex\hbox{$/$}\kern-.53em\hbox{$v$}}
\newcommand{\slF}{\raise.15ex\hbox{$/$}\kern-.53em\hbox{$F$}}
\newcommand{\slL}{\raise.15ex\hbox{$/$}\kern-.53em\hbox{$L$}}
\newcommand{\slP}{\raise.15ex\hbox{$/$}\kern-.53em\hbox{$P$}}
\newcommand{\slp}{\raise.15ex\hbox{$/$}\kern-.53em\hbox{$p$}}
\newcommand{\slq}{\raise.15ex\hbox{$/$}\kern-.53em\hbox{$q$}}
\newcommand{\slR}{\raise.15ex\hbox{$/$}\kern-.53em\hbox{$R$}}
\newcommand{\slQ}{\raise.15ex\hbox{$/$}\kern-.53em\hbox{$Q$}}
\newcommand{\slK}{\raise.15ex\hbox{$/$}\kern-.53em\hbox{$K$}}
\newcommand{\slk}{\raise.15ex\hbox{$/$}\kern-.53em\hbox{$k$}}
\newcommand{\slD}{\raise.15ex\hbox{$/$}\kern-.73em\hbox{$D$}}
\newcommand{\slC}{\raise.15ex\hbox{$/$}\kern-.53em\hbox{$C$}}
\newcommand{\slA}{\raise.15ex\hbox{$/$}\kern-.53em\hbox{$A$}}
\newcommand{\slSigma}{\raise.15ex\hbox{$/$}\kern-.53em\hbox{$\Sigma$}}
\newcommand{\slpartial}{\raise.15ex\hbox{$/$}\kern-.53em\hbox{$\partial$}}
\newcommand{\slcalP}{\raise.15ex\hbox{$/$}\kern-.63em\hbox{$\cal P$}}

\def\p{{\boldsymbol p}}
\def\P{{\boldsymbol P}}

\def\l{{\boldsymbol l}}
\def\k{{\boldsymbol k}}

\def\x{{\boldsymbol x}}
\def\y{{\boldsymbol y}}

\def\v{{\boldsymbol v}}

\def\u{{\boldsymbol u}}
\def\rmd{{\rm d}}

\begin{document}

\title{\bf %
Non-renormalizability  of the\\ classical statistical approximation}
\author{Thomas Epelbaum, Fran\c cois Gelis, Bin Wu}
\maketitle
 \begin{center}
  Institut de Physique Th\'eorique (URA 2306 du CNRS)\\
   CEA/DSM/Saclay, 91191 Gif-sur-Yvette Cedex, France
 \end{center}

\begin{abstract}
  In this paper, we discuss questions related to the renormalizability
  of the classical statistical approximation, an approximation scheme
  that has been used recently in several studies of out-of-equilibrium
  problems in Quantum Field Theory. Although the ultraviolet power
  counting in this approximation scheme is identical to that of the
  unapproximated quantum field theory, this approximation is not
  renormalizable. The leading cause of this non-renormalizability is
  the breakdown of Weinberg's theorem in this approximation. We also
  discuss some practical implications of this negative result for
  simulations that employ this approximation scheme, and we speculate
  about a possible modification of the classical statistical
  approximation in order to systematically subtract the leading
  residual divergences.
\end{abstract}

\section{Introduction}
\label{sec:intro}
In recent years, there has been a lot of interest in the study of
out-of-equilibrium systems in quantum field theory, in view of
applications to high energy heavy ion collisions, cosmology, or cold
atom
physics~\cite{BergeBG1,BergeBS1,BergeBSV1,BergeBSV2,BergeBSV3,BergeGSS1,BergeS4,BergeSS3,FukusG1,Fukus3,KunihMOST1,IidaKMOS1,DusliEGV1,EpelbG1,DusliEGV2,EpelbG2,EpelbG3,KurkeM1,KurkeM2,KurkeM3,YorkKLM1,AttemRS1,BlaizGLMV1,BlaizLM2,HelleJW1,CasalHMS1,Wu1,ReinoS1,GautiS1,GirauS1,FloerW1,GasenP1,GasenKP1}. Generically,
the question one would like to address is that of a system prepared in
some non-equilibrium initial state, and let to evolve under the sole
self-interactions of its constituents.  The physically relevant
quantum field theories cannot be solved exactly, and therefore some
approximation scheme is mandatory in order to make progress. Moreover,
the standard perturbative expansion in powers of the interaction
strength is in general ill-suited to these out-of-equilibrium
problems. Indeed, the coefficients in the perturbative expansion are
time dependent and generically growing with time, thereby voiding the
validity of the expansion after some finite time.

This ``secularity'' problem is resolved by the resummation of an
infinite set of perturbative contributions, which can be achieved via
several schemes. The simplest of these schemes is kinetic
theory. However, in order to obtain a Boltzmann equation from the
underlying quantum field theory, several important assumptions are
necessary~\cite{ArnolMY5}: (i) a relatively smooth system, so that a gradient
expansion can be performed, and (ii) the existence of well-defined
quasiparticles. These limitations, especially the latter, make kinetic
theory difficult to justify for describing the early stages of heavy
ion collisions.

Closer to the underlying quantum field theory, two resummation schemes
have been widely considered in many works. One of them is the
2-particle irreducible (2PI)
approximation~\cite{LuttiW1,BaymK1,DominM1,DominM2,CornwJT1,Berge3,BergeBRS1,ReinoS2,BransG1,AartsLT1,ReinoS1,HattaN2,HattaN1}. This
scheme consists in solving the Dyson-Schwinger equations for the
2-point functions (and possibly for the expectation value of the
field, if it differs from zero). The self-energy diagrams that are
resummed on the propagator are obtained self-consistently from the sum
of 2PI skeleton vacuum diagrams (often denoted $\Gamma_2[G]$ in the
literature). The only approximation arises from the practical
necessity of truncating the functional $\Gamma_2[G]$ in order to have
manageable expressions. In applications, the 2PI scheme suffers from
two limitations. One of them is purely computational: the convolution
of the self-energy with the propagator takes the form of a memory
integral, that in principle requires that one stores the entire
history of the evolution of the system, from the initial time to the
current time. The needed storage therefore grows quadratically with
time\footnote{This can be alleviated somewhat by an extra
  approximation, in which one stores only the ``recent'' history of
  the system, in a sliding time window that moves with the current
  time.}. The second difficulty appears in systems that are the siege
of large fields, or large occupation numbers. For instance, in QCD,
``large'' would mean of order $g^{-1}$ for the fields, and of order
$g^{-2}$ for the gluon occupation number. In this regime, the
functional $\Gamma_2[G]$ contains terms that have the same order of
magnitude at every order in the loop expansion, and therefore one
cannot justify to truncate it at a finite loop order\footnote{Such a
  truncation becomes legitimate, even in the large field regime, if
  there is an additional expansion parameter that one can use to
  control the loop expansion in $\Gamma_2[G]$. In some theories with a
  large number $N$ of constituents (e.g. an $O(N)$ scalar theory in
  the limit $N\to\infty$), one can compute exactly the leading term of
  $\Gamma_2[G]$ in the $1/N$ expansion~\cite{Berge3}.}.  It turns out
that these problems of strong fields occur in real world problems,
e.g. in the early stages of heavy ion
collisions~\cite{IancuLM3,IancuV1,LappiM1,GelisIJV1,Gelis15}.

There is an alternative resummation scheme, that includes all the
leading contributions in the large field regime and is similarly free
of secular terms, called the Classical Statistical Approximation
(CSA)~\cite{PolarS1,Son1,KhlebT1,GelisLV2,FukusGM1,Jeon4}. It owes its
name to the way it is implemented in practice, as an average over
classical solutions of the field equations of motion, with a Gaussian
statistical ensemble of initial conditions. The ability of this method
to remain valid in the large field regime comes with a tradeoff~: the
CSA can be tuned to be exact at the 1-loop level, but starting at the
2-loop order and beyond, it includes only a subset of all the possible
contributions.  The CSA can be derived via several methods~: from the
path integral representation of observables~\cite{Jeon4}, as an
approximation at the level of the diagrammatic rules in the
retarded-advanced formalism, or as an exponentiation of the 1-loop
result~\cite{GelisLV2}.

The diagrammatic rules that define the classical statistical
approximation allow graphs that have arbitrarily many loops.  As in
any field theory, the loops that arise in this expansion involve an
integral over a 4-momentum, and this integral can be ultraviolet
divergent. In the underlying --non approximated-- field theory, we
know how to deal with these infinities by redefining a finite number
of parameters of the Lagrangian (namely the coupling constant, the
mass and the field normalization). In general, this is done by first
introducing an ultraviolet regulator, for instance a momentum cutoff
$\Lambda_{_{\rm UV}}$ on the loop momenta, and by letting the bare
parameters of the Lagrangian depend on $\Lambda_{_{\rm UV}}$ in such a way
that physical quantities are independent of $\Lambda_{_{\rm UV}}$ (and of
course are finite in the limit $\Lambda_{_{\rm UV}}\to \infty$). That this
redefinition is possible is what characterizes a renormalizable field
theory. 

In contrast, non-renormalizable theories are theories in which one
needs to introduce new operators that did not exist in the Lagrangian
one started from, in order to subtract all the ultraviolet divergences
that arise in the loop expansion. This procedure defines an
``ultraviolet completion'' of the original theory, which is well
defined at arbitrary energy scales. The predictive power of the
original theory is limited by the order at which it becomes necessary
to introduce these new operators\footnote{The predictive power of its
  ultraviolet completion may be quite limited as well, depending on
  how many new operators need to be introduced at each order
  (especially if this number grows very quickly or even worse becomes
  infinite).}.

It can also happen that, starting from a renormalizable field theory,
certain approximations of this theory (for instance including certain
loop corrections, but not all of them) are not renormalizable. This
will be our main concern in this paper, in the context of the
classical statistical approximation. A recent numerical study
\cite{BergeBSV3} showed a pronounced cutoff dependence for rather
large couplings in a computation performed in this approximation
scheme. This could either mean that the CSA is not renormalizable, or
that the CSA is renormalizable but that renormalization was not
performed properly in this computation. It is therefore of utmost
importance to determine to which class --re\-nor\-ma\-li\-za\-ble or
non-renormalizable-- the CSA belongs, since this has far reaching
practical implications on how it can be used in order to make
predictive calculations, and how to interpret the existing
computations.

Note that the question of the renormalizability of classical
approximation schemes has already been discussed in quantum field
theory at finite temperature
\cite{BodekMS1,ArnolSY1,AartsS1,AartsS3,AartsNW1}, following attempts
to calculate non-perturbatively the sphaleron transition rate
\cite{GrigoR1,GrigoRS1,GrigoRS2,AmbjoLS1,AmbjoLS2,AmbjoAPS2,AmbjoAPS1,AmbjoF1}.
In this context, one is calculating the leading high temperature
contribution, and in the classical approximation the Bose-Einstein
distribution gets replaced by $T/\omega_\k$. This approximation leads
to ultraviolet divergences in thermal contributions, that would
otherwise be finite thanks to the exponential tail of the
Bose-Einstein distribution. However, it has been shown that only a
finite number of graphs have such divergences, and that they can all
be removed by appropriate counterterms. The problem we will consider
in this paper is different since we are interested in the classical
approximation of a zero-temperature quantum field theory, where the
factors $T/\omega_\k$ are replaced by $1/2$. This changes drastically
the ultraviolet behavior.

In the section \ref{sec:prelim}, we expose the scalar toy model we are
going to use throughout the paper as a support of this discussion, we
also remind the reader of the closed time path formalism and of the
retarded-advanced formalism (obtained from the latter via a simple
field redefinition), and we present the classical statistical
approximation in two different ways (one that highlights its
diagrammatic rules, and one that is more closely related to the way it
is implemented in numerical simulations). Then, we analyze in the
section \ref{sec:UVcount} the ultraviolet power counting in the CSA,
and show that it is identical to that in the underlying field theory.
In the section \ref{sec:nonren}, we examine all the one-loop 2-point
and 4-point functions in the CSA, and we show that one of them
violates Weinberg's theorem. This leads to contributions that are
non-renormalizable in the CSA. In the section \ref{sec:int}, we
discuss the implications of non-renormalizability of the CSA for the
calculation of some observables. We also argue that it may be possible
to systematically subtract the leading non-renormalizable terms by the
addition of a complex noise term to the classical equations of motion.
Finally, the section \ref{sec:concl} is devoted to concluding remarks.
Some technical derivations are relegated into two appendices.

\section{Preliminaries}
\label{sec:prelim}
\subsection{Toy model}
\label{sec:model}
In order to illustrate our point, let us consider a massless real
scalar field $\phi$ in four space-time dimensions, with quartic
self-coupling, and coupled to an external source $j(x)$,
\begin{equation}
{\cal L}\equiv \frac{1}{2}(\partial_\mu\phi)(\partial^\mu\phi)-\frac{m^2}{2}\phi^2
-\frac{g^2}{4!}\phi^4+j\phi\; .
\end{equation}
In this model, $j(x)$ is a real valued function, given once for all as
a part of the description of the model. Sufficient regularity and
compactness of this function will be assumed as necessary.

We also assume that the state of the system at $x^0=-\infty$ is the
vacuum state (by adiabatically turning off the couplings at asymptotic
times, we can assume that this is the perturbative vacuum state
$\big|0_{\rm in}\big>$). Because of the coupling to the external
source $j(x)$, the system is driven away from the vacuum state, and
observables measured at later times acquire non-trivial values.  Our
goal is to compute the expectation value of such observables,
expressed in terms of the field operator and its derivatives, in the
course of the evolution of the system,
\begin{equation}
\left<{\cal O}\right>\equiv \big<0_{\rm in}\big|{\cal O}[\phi,\partial\phi]\big|0_{\rm in}\big>\; .
\label{eq:obs0}
\end{equation}
For simplicity, one may assume that the observable is a local
(i.e. depends on the field operator at a single space-time point) or
multi-local operator (i.e. depends on the field operator at a finite
set of space-time points).

\subsection{Closed time path formalism}
\label{sec:CTP}
It is well known that the proper framework to compute expectation
values such as the one defined in eq.~(\ref{eq:obs0}) is the
Schwinger-Keldysh (or ``closed time path'')
formalism~\cite{Schwi1,Keldy1}. In this formalism, there are two
copies $\phi_+$ and $\phi_-$ of the field (corresponding respectively
to fields in amplitudes and fields in complex conjugated amplitudes),
and four bare propagators depending on which type of fields they
connect. The expectation value of eq.~(\ref{eq:obs0}) can be expanded
diagrammatically (each loop brings an extra power of the coupling
$g^2$) by a set of rules that generalize the traditional Feynman rules
in a simple manner~:
\begin{itemize}
\item[{\bf i.}] Each vertex of a graph can be of type $+$ or $-$, and
  for a given graph topology one must sum over all the possible
  assignments of the types of these vertices. The rule for the $+$
  vertex ($-ig^2$) and for the $-$ vertex ($+ig^2$) differ only in
  their sign. The same rule applies to the external source $j$.
\item[{\bf ii.}] A vertex of type $\epsilon$ and a vertex of type
  $\epsilon'$ must be connected by a bare propagator
  $G^0_{\epsilon\epsilon'}$. In momentum space, these bare propagators
  read~:
  \begin{align}
    &G^0_{++}(p)=\frac{i}{p^2-m^2+i\epsilon}\;,&&G^0_{--}(p)=\frac{-i}{p^2-m^2-i\epsilon}\nonumber\\
    &G^0_{+-}(p)=2\pi\theta(-p^0)\delta(p^2-m^2)\;,&&G^0_{-+}(p)=2\pi\theta(p^0)\delta(p^2-m^2)
    \label{eq:SKprops}
  \end{align}
\end{itemize}
The four bare propagators of the Schwinger-Keldysh formalism are
related by a simple algebraic identity,
\begin{equation}
  G^0_{++}+G^0_{--}=G^0_{+-}+G^0_{-+}\; ,
  \label{eq:SKident}
\end{equation}
that one can check immediately from eqs.~(\ref{eq:SKprops}). Note
that, on a more fundamental level, this identity follows from the
definition of the various $G_{\epsilon\epsilon'}$ as vacuum
expectation values of pairs of fields ordered in various ways. For
this reason, it is true not only for the bare propagators, but for
their corrections at any order in $g^2$.

\subsection{Retarded-advanced formalism}
\label{sec:RA}
The Schwinger-Keldysh formalism is not the only one that can be used
to calculate eq.~(\ref{eq:obs0}). One can arrange the four bare
propagators $G^0_{\epsilon\epsilon'}$ in a $2\times 2$ matrix, and
obtain equivalent diagrammatic rules by applying a ``rotation'' to
this matrix~\cite{AurenB1,EijckKW1,Gelis3,AartsS3}. Among this family of
transformations, especially interesting are those that exploit the
linear relationship (\ref{eq:SKident}) among the
$G^0_{\epsilon\epsilon'}$ in order to obtain a vanishing entry in the
rotated matrix. The retarded-advanced formalism belongs to this class
of transformations, and its propagators are defined by (let us denote
$\alpha=1,2$ the two values taken by the new index)~:
\begin{eqnarray}
{\mathbbm G}_{\alpha\beta}^0\equiv 
\sum_{\epsilon,\epsilon^\prime=\pm}
\Omega_{\alpha\epsilon}\Omega_{\beta\epsilon^\prime}
{G}_{\epsilon\epsilon^\prime}^0\; ,
\label{eq:SK-rotation}
\end{eqnarray}
with the transformation matrix defined as 
\begin{equation}
\Omega_{\alpha\epsilon}
\equiv
\begin{pmatrix}
1 & -1 \\
1/2 & 1/2 \\
\end{pmatrix}\; .
\end{equation}
The bare rotated propagators read
\begin{equation}
{\mathbbm G}_{\alpha\beta}^0
=
\begin{pmatrix}
0 & G_{_A}^0\\
G_{_R}^0& G_{_S}^0\\
\end{pmatrix}\; ,
\end{equation}
where we have introduced
\index{Retarded propagator}
\begin{equation}
G_{_R}^0 = G_{++}^0-G_{+-}^0\;,\;
G_{_A}^0 = G_{++}^0-G_{-+}^0\;,\;
G_{_S}^0 = \frac{1}{2}(G_{++}^0+G_{--}^0)\; .
\end{equation}
(The subscripts R, A and S stand respectively for {\sl retarded}, {\sl
  advanced} and {\sl symmetric}.)

It is straightforward to verify that in the rotated formalism, the
various vertices read~:
\begin{equation}
\Gamma_{\alpha\beta\gamma\delta}\equiv -ig^2\left[
  \Omega^{-1}_{+\alpha}\Omega^{-1}_{+\beta}\Omega^{-1}_{+\gamma}\Omega^{-1}_{+\delta}
  -
  \Omega^{-1}_{-\alpha}\Omega^{-1}_{-\beta}\Omega^{-1}_{-\gamma}\Omega^{-1}_{-\delta}
\right]\; ,
\end{equation}
where
\begin{equation}
\Omega^{-1}_{\epsilon\alpha}=
\begin{pmatrix}
1/2 & 1 \\
-1/2 & 1 \\
\end{pmatrix}\qquad\qquad[\Omega_{\alpha\epsilon}\Omega_{\epsilon\beta}^{-1}=\delta_{\alpha\beta}]\; .
\end{equation}
More explicitly, we have~:
\begin{eqnarray}
&&\Gamma_{1111}=\Gamma_{1122}=\Gamma_{2222}=0\nonumber\\
&&\Gamma_{1222}=-ig^2\; ,\quad \Gamma_{1112}=-ig^2/4\; .
\end{eqnarray}
(The vertices not listed explicitly here are obtained by trivial
permutations.)  Concerning the insertions of the external source, the
diagrammatic rules in the retarded-advanced formalism are~:
\begin{equation}
J_1=ij\;,\quad J_2 = 0\; .
\end{equation}

\subsection{From the external source to an external classical field}
\label{sec:extfield}
\begin{figure}[htbp]
\begin{center}
\setbox1\hbox to 9cm{\hfil\resizebox*{9cm}{!}{\includegraphics{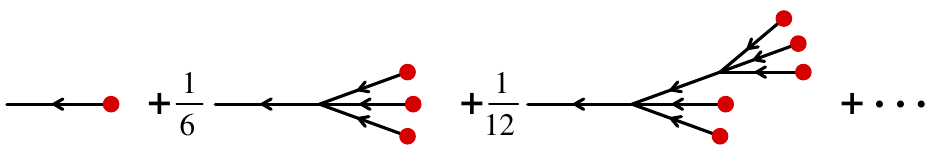}}}
$\varphi\equiv\raise -4.3mm\box1$
\end{center}
\caption{\label{fig:phi4-tree}The first three terms of the
  diagrammatic expansion of the external field in terms of the
  external source, for a field theory with quartic coupling. The red
  dots represent the source insertions $J_1$, and the lines with an
  arrow are bare retarded propagators $G_{21}^0$. The quartic vertices
  are all of the type $\Gamma_{1222}$.}
\end{figure}

From the above rules, we see that an external source can only be
attached to a propagator endpoint of type 1, i.e. to the lowest time
endpoint of a retarded or advanced propagator ($G_{12}^0=G_{_A}^0$,
$G_{21}^0=G_{_R}^0$), as in the formula
\begin{equation}
\int \rmd^4y\; G_{21}^0(x,y)\,J_1(y)\; .
\end{equation}
(This expression corresponds to the first graph on the left of the
figure \ref{fig:phi4-tree}.) It is easy to see that the external
source can be summed to all orders, if one introduces the object
$\varphi(x)$ defined diagrammatically in the figure \ref{fig:phi4-tree}.
It is well known that this series obeys the classical equation of motion,
\begin{equation}
(\square+m^2)\varphi + \frac{g^2}{6}\varphi^3=j\; ,
\end{equation}
and since all the propagators are all of type $G_{21}^0$, i.e. retarded,
it obeys the following boundary condition~:
\begin{equation}
\lim_{x^0\to-\infty}\varphi(x)=0\; .
\end{equation}
The source $j$ can be completely eliminated from the diagrammatic
rules, by adding to the Lagrangian couplings between the field
operator $\phi$ and the classical field $\varphi$,
\begin{equation}
\Delta{\cal L}\equiv 
g^2\left[
\frac{1}{2}\;\varphi^2\;\phi_1\;\phi_2
+
\frac{1}{2}\;\varphi\; \phi_1\;\phi_2^2
+
\frac{1}{4!}\;\varphi\; \phi_1^3
\right]
\; .
\label{eq:dL}
\end{equation}
Note that, since in all the graphs in the figure \ref{fig:phi4-tree}
the root of the tree is terminated by an index $2$, the classical
field $\varphi$ can only be attached to an index of type $2$ in these vertices.

\subsection{Classical statistical approximation (CSA)}
\label{sec:CSAdef}
\subsubsection{Definition from truncated retarded-advanced rules}
\label{sec:CSAdef1}
The classical statistical approximation consists in dropping all the
graphs that contain the vertex $\Gamma_{2111}$, i.e. in assuming~:
\begin{equation}
\Gamma_{2111}=0\qquad (\mbox{and similarly for the permutations of $2111$})\; .
\end{equation}
In the rest of this paper, we will simply call CSA the field theory
obtained by dropping all the vertices that have 3 indices of type 1 in
the retarded-advanced formalism, while everything else remains
unchanged. Therefore, in the retarded-advanced formalism, the CSA is
defined by the following diagrammatic rules~:
\begin{itemize}
  \item[{\bf i.}]{\bf bare propagators~:}
    \begin{eqnarray}
      &&
      G_{21}^0(p) = \frac{i}{(p^0+i\epsilon)^2-\p^2-m^2}\;,\quad
      G_{12}^0(p) = \frac{i}{(p^0-i\epsilon)^2-\p^2-m^2}\;,\nonumber\\
      &&
      G_{22}^0(p) = \pi\,\delta(p^2-m^2)\;.
      \label{eq:RA:prop}
    \end{eqnarray}
  \item[{\bf ii.}]{\bf vertices~:}
    \begin{equation}
      \Gamma_{1222}\mbox{\ (and permutations)}=-ig^2\quad,\quad\mbox{all other combinations zero}\; .
      \label{eq:RA:vtx}
    \end{equation}
    \item[{\bf iii.}]{\bf external sources~:}
      \begin{equation}
        J_1=ij\;,\quad J_2 = 0\; .
        \label{eq:RA:J}
      \end{equation}
    \item[{\bf iii${}^\prime$.}]{\bf external field~(see the section \ref{sec:extfield})~:}
      \begin{equation}
        \Phi_1=0\;,\quad \Phi_2 = \varphi\; .
        \label{eq:RA:phi}
      \end{equation}
\end{itemize}
Note that this ``truncated'' field theory is still quite non trivial,
in the sense that the above diagrammatic rules allow graphs with
arbitrarily many loops.  The numerical simulations that implement the
CSA provide the sum to all orders of the graphs that can be
constructed with these rules (with an accuracy in principle only
limited by the statistical errors in the Gaussian average over the
initial field fluctuations, since this average is approximated by a
Monte-Carlo sampling).

\subsubsection{Definition by exponentiation of the 1-loop result}
\label{sec:CSAdef2}
The previous definition of the CSA makes it very clear what graphs are
included in this approximation and what graphs are not. However, it is
a bit remote from the actual numerical implementation. Let us also
present here an alternative --but strictly equivalent-- way of
introducing the classical statistical approximation, that directly
provides a formulation that can be implemented numerically.

Firstly, observables at leading order are expressible in terms of the
retarded classical field $\varphi$ introduced above,
\begin{equation}
\big<0_{\rm in}\big|{\cal O}[\phi,\partial\phi]\big|0_{\rm in}\big>_{\rm LO}
=
{\cal O}[\varphi,\partial\varphi]\; .
\label{eq:LO}
\end{equation}
At next-to-leading order, it has been shown in
\cite{GelisLV2,GelisLV3,GelisLV4} that the observable can be expressed
as follows,
\begin{eqnarray}
&&\big<0_{\rm in}\big|{\cal O}[\phi,\partial\phi]\big|0_{\rm in}\big>_{\rm NLO}
=
\nonumber\\
&&\qquad\quad=
\left[\frac{1}{2}\int\frac{\rmd^3\k}{(2\pi)^3 2E_\k}
\int \rmd^3\u\, \rmd^3\v\,
({\bs\alpha}_\k\cdot{\mathbbm T}_\u)
({\bs\alpha}_\k^*\cdot{\mathbbm T}_\v)
\right]\,
{\cal O}[\varphi,\partial\varphi]\, .
\label{eq:NLO}
\end{eqnarray}
In eq.~(\ref{eq:NLO}), the operator ${\mathbbm T}_\u$ is the generator
of shifts of the initial condition for the classical field $\varphi$
on some constant time surface (the integration surface for the
variables $\u$ and $\v$) located somewhere before the source $j$ is
turned on\footnote{This is why eq.~(\ref{eq:NLO}) does not have a term
  linear in ${\mathbbm T}_\u$, contrary to the slightly more general
  formulas derived in refs.~\cite{GelisLV2,GelisLV3,GelisLV4}.}. This
means that if we denote $\varphi[\varphi_{\rm init}]$ the classical
field as a functional of its initial condition, then for any
functional $F[\varphi]$ of $\varphi$, we have
\begin{equation}
\left[\exp\int \rmd^3\u\;(a\cdot{\mathbbm T}_\u)\right]\;F[\varphi[\varphi_{\rm init}]]=F[\varphi[\varphi_{\rm init}+a]]\; .
\end{equation}
(This equation can be taken as the definition of ${\mathbbm T}_\u$.)
In eq.~(\ref{eq:NLO}), the fields ${\bs\alpha}_\k$ are free plane waves of momentum $k^\mu$~:
\begin{equation}
{\bs\alpha}_\k(u)\equiv e^{ik\cdot u}\;,\qquad (\square+m^2){\bs\alpha_k}(u) =0
\; .
\end{equation}
Note that in eq.~(\ref{eq:NLO}), the integration variable $\k$ is a
loop momentum. In general, the integral over $\k$ therefore diverges
in the ultraviolet, and must be regularized by a cutoff. After the
Lagrangian parameters have been renormalized at 1-loop, this cutoff
can be safely sent to infinity.

In this framework, the classical statistical method is defined as the
result of the exponentiation of the operator that appears in the right
hand side of eq.~(\ref{eq:NLO}),
\begin{eqnarray}
&&
\big<0_{\rm in}\big|{\cal O}[\phi,\partial\phi]\big|0_{\rm in}\big>_{\rm CSA}
=\nonumber\\
&&\qquad=
\exp\left[\frac{1}{2}\int\!\!\frac{\rmd^3\k}{(2\pi)^3 2E_\k}
\int\!\! \rmd^3\u\, \rmd^3\v\,
({\bs\alpha}_\k\cdot{\mathbbm T}_\u)
({\bs\alpha}_\k^*\cdot{\mathbbm T}_\v)
\right]
{\cal O}[\varphi,\partial\varphi]\, .
\label{eq:CSA}
\end{eqnarray}
Note that by construction, the CSA is identical to the underlying
theory at LO and NLO, and starts differing from it at
NNLO and beyond (some higher loop graphs are included but not all
of them). The relation between this formula and the way the classical
statistical method is implemented lies in the fact that the
exponential operator is equivalent to a Gaussian average over a
Gaussian distribution of initial conditions for the classical field
$\varphi$,
\begin{eqnarray}
&&\exp\left[\frac{1}{2}\int\frac{\rmd^3\k}{(2\pi)^3 2E_\k}
\int \rmd^3\u\, \rmd^3\v\,
({\bs\alpha}_\k\cdot{\mathbbm T}_\u)
({\bs\alpha}_\k^*\cdot{\mathbbm T}_\v)
\right]\;
F[\varphi[\varphi_{\rm init}]]
\nonumber\\
&&\qquad\qquad\qquad\qquad
=
\int[{\rm D}a(\u){\rm D}\dot{a}(\u)]
\;G[a,\dot{a}]\;F[\varphi[\varphi_{\rm init}+a]]\; ,
\label{eq:CSA1}
\end{eqnarray}
where $G[a,\dot{a}]$ is a Gaussian distribution, whose elements can be
generated as
\begin{equation}
a(u)=\int\frac{\rmd^3\k}{(2\pi)^3 2E_\k}\;\left[
c_\k\,{\bs\alpha}_\k(u)+c_\k^*\,{\bs\alpha}_\k^*(u)
\right]\; ,
\label{eq:fluct}
\end{equation}
with $c_\k$ complex Gaussian random numbers defined by
\begin{equation}
\big<c_\k\big>=0\quad,\quad \big<c_\k c_\l\big>=0\quad,\quad 
\big<c_\k c_\l^*\big>=(2\pi)^3 E_\k\delta(\k-\l)\; .
\end{equation}
We will not show here the equivalence of the two ways of defining the
classical statistical approximation that we have exposed in this
section. The main reason for recalling the second definition of the
CSA was to emphasize the meaning of the variable $\k$ in
eqs.~(\ref{eq:NLO}), (\ref{eq:CSA}) and (\ref{eq:fluct}), as a loop
momentum. Therefore, an upper limit introduced in the $\k$-integration
in any of these formulas will effectively play the role of an
ultraviolet cutoff that regularizes loop integrals.

To make the connection with the diagrammatic rules of the classical
statistical approximation introduced in the previous subsection, the
cutoff on the momentum of the initial fluctuations is an upper limit
for the momentum flowing through the $G_{22}^0$ propagators. In
contrast, the largest momentum that can flow through the $G_{21}^0$
and $G_{12}^0$ propagators is only controlled by the discretization of
space, i.e. by the inverse lattice cutoff. In some implementations,
these two cutoffs are identical, but other implementations have chosen
to have distinct cutoffs for these two purposes\footnote{The downside
  of having two separate cutoffs is that this form of regularization
  violates~\cite{EpelbGTW1} the Kubo-Martin-Schwinger
  identities~\cite{Kubo1,MartiS1}, which leads to a non-zero
  scattering rate even in the vacuum.}~:
\begin{itemize}
\item In Refs.~\cite{BergeBS1,BergeBSV1,BergeBSV2} an explicit cutoff
  $\Lambda_{_{\rm UV}}$, distinct from the lattice cutoff, is
  introduced in order to limit the largest $\k$ of the initial
  fluctuations. In this setup, $\Lambda_{_{\rm UV}}$ is smaller than
  the lattice momentum cutoff, and the lattice spacing no longer
  controls the ultraviolet limit of the computation.
\item In Refs.~\cite{DusliEGV1,EpelbG1,DusliEGV2,EpelbG3,BergeBSV3}
  fluctuation modes are included up to the lattice momentum cutoff,
  i.e. $\Lambda_{_{\rm UV}}$ is inversely proportional to the lattice
  spacing.
\end{itemize}
A common caveat of most of these computations is that none has studied
the behavior of the results in the limit $\Lambda_{_{\rm UV}}\to
\infty$, at the exception of ref.~\cite{BergeBSV3} where a strong
dependence on the ultraviolet cutoff was found.

\section{Ultraviolet power counting}
\label{sec:UVcount}
After having defined the classical statistical approximation, we can
first calculate the superficial degree of ultraviolet divergence for
arbitrary graphs in the CSA, in order to see what kind of divergences
one may expect. This is best done by using the definition introduced
in the section \ref{sec:CSAdef1}, that defines the CSA by its
diagrammatic rules.

Let us consider a generic {\sl connected} graph ${\cal G}$ built with
these diagrammatic rules, made of~:
\begin{itemize}
  \item $E$ external legs
  \item $I$ internal lines
  \item $L$ independent loops
  \item $V$  vertices  of type $\phi^4$
  \item $V_2$ vertices  of type $\phi^2\varphi^2$
  \item $V_1$ vertices  of type $\phi^3\varphi$
\end{itemize}
Note that for the internal lines, the superficial degree of divergence
does not distinguish\footnote{As we shall see later, due to the
  peculiar analytic structure of the integrands of graphs in the CSA,
  this power counting is too naive to accurately reflect the actual
  ultraviolet divergences.} between the propagators $G_{12}^0$,
$G_{21}^0$ and $G_{22}^0$, because they all have a mass dimension
$-2$. These numbers are related by the following relations~:
\begin{eqnarray}
  E+2I &=& 4V+3V_1+2V_2\;,\\
  L&=& I - (V+V_1+V_2) +1\; .
\end{eqnarray}
The first of these identities states that the number of propagator
endpoints must be equal to the number of slots where they can be
attached to vertices. The second identity counts the number of
independent momenta that can circulate in the loops of the graph.

In terms of these quantities, the superficial degree of divergence of
the graph ${\cal G}$ is given by
\begin{eqnarray}
\omega({\cal G})&=& 4L-2I\nonumber\\
&=& 4-E-(V_1+2V_2)\nonumber\\
&=& 4-E-N_\varphi\; ,
\label{eq:UVcount}
\end{eqnarray}
where $N_\varphi\equiv V_1+2V_2$ is the number of powers of the
external classical field $\varphi$ inserted into the graph ${\cal G}$.

Note that $4-E$ is the superficial degree of divergence of a graph
with $E$ external points in a 4-dimensional scalar $\phi^4$ theory, in
the absence of an external field/source. Therefore, the external field
can only decrease the superficial degree of divergence (since
$N_\varphi\ge 0$), which was expected since the couplings to the
external field (see eq.~(\ref{eq:dL})) have a positive mass dimension,
i.e. they are super-renormalizable interactions.

The crucial point about this formula is that the superficial degree of
divergence does not depend on the fact that we have excluded the
vertices of type $2111$ in the classical statistical approximation. In
other words, the ultraviolet power counting is exactly the same in the
full theory and in the CSA. Eq.~(\ref{eq:UVcount}) suggests that the
only ultraviolet divergent quantities are those for which $E\le 4$,
exactly as in the unapproximated theory.

As we shall see in the next section, there is nevertheless an issue
that hinders the renormalizability of the classical statistical
approximation. Discarding the $\Gamma_{1112}$ alters in subtle ways
the analytic structure of Green's functions, which leads to a
violation of Weinberg's theorem\footnote{In addition to power counting
  arguments, renormalizability requires some handle on the recursive
  structure of the ultraviolet divergences. This may come in the form
  of Dyson's convergence theorem \cite{Dyson2}, whose proof was
  completed by Weinberg \cite{Weinb4} and somewhat simplified by Hahn
  and Zimmermann \cite{HahnZ1,Zimme1} (see also
  Refs.~\cite{CasweK1,CasweK2}). This result states that if all the
  divergences in the subgraphs of a given graph ${\cal G}$ have been
  subtracted, then the remaining divergence is a polynomial of degree
  $\omega({\cal G})$ in the external momenta.}. As a consequence,
ultraviolet divergences in the CSA can be stronger that one would
expect on the basis of the power counting alone.

\section{Ultraviolet divergences in the CSA}
\label{sec:nonren}
\subsection{Introduction}
The un-truncated $\phi^4$ theory that we started from is well known to
be renormalizable\footnote{The fact that we are dealing here with a
  field theory coupled to an external source does not spoil this
  property, for sufficiently smooth external sources. See
  Ref.~\cite{Colli1}, chapter 11.}. This means that all its
ultraviolet divergences can be disposed of by redefining the
coefficients in front of the operators that appear in the bare
Lagrangian.

In the retarded-advanced basis, the Lagrangian of the CSA differs from
the Lagrangian of the unapproximated theory in the fact that the vertex
$\Gamma_{1112}$ is missing. All the other terms of the Lagrangian are
unchanged, in particular the operators that are quadratic in the
fields. In this section we systematically examine 2- and 4-point
functions at one loop, in order to see whether their ultraviolet
behavior is compatible with renormalizability or not.

\subsection{Self-energies at one loop}
Let us start with the simplest possible loop correction: the one-loop
self-energy, made of a tadpole graph. Depending on the indices $1$ and
$2$ assigned to the two external legs, these self-energies are given
in eq.~(\ref{eq:S-1loop}).
\setbox2\hbox to 1.2cm{\resizebox*{1.2cm}{!}{\includegraphics{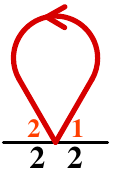}}}
\setbox3\hbox to 1.2cm{\resizebox*{1.2cm}{!}{\includegraphics{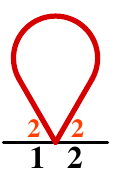}}}
\begin{eqnarray}
-i\big[\Sigma_{11}\big]_{_{\rm CSA}}^{\rm 1\ loop}&=&0
\quad,\qquad
-i\big[\Sigma_{22}\big]_{_{\rm CSA}}^{\rm 1\ loop}=\raise -2.5mm\box2=0
\nonumber\\
-i\big[\Sigma_{12}\big]_{_{\rm CSA}}^{\rm 1\ loop}&=&\raise -2.5mm\box3=-i\frac{g^2\Lambda_{_{\rm UV}}^2}{16\pi^2}\; .
\label{eq:S-1loop}
\end{eqnarray}
$\Sigma_{11}$ is zero at one loop in the CSA, because it requires a
vertex $1112$ that has been discarded. $\Sigma_{22}$ is also zero,
because it contains a closed loop made of a retarded propagator.  The
only non-zero self-energy at 1-loop is $\Sigma_{12}$, that displays
the usual quadratic divergence. This can be removed by a mass
counterterm in the Lagrangian,
\begin{equation}
\delta m^2 = -\frac{g^2\Lambda_{_{\rm UV}}^2}{16\pi^2}\; ,
\end{equation}
since the mass term in the Lagrangian is precisely a $\phi_1\phi_2$ operator.

\subsection{Four point functions at one loop}
\subsubsection{Vanishing functions~: $\Gamma_{1112}$, $\Gamma_{1111}$ and $\Gamma_{2222}$}
The 4-point function with indices $1112$ is a prime suspect for
Green's functions that may cause problems with the renormalizability
of the CSA. Indeed, the CSA consists in discarding the operator
corresponding to this vertex from the Lagrangian. Therefore, if an
intrinsic\footnote{Here, we are talking about the overall divergence
  of the function, not the divergences associated to its various
  subgraphs, that may be subtracted by having renormalized the other
  operators of the Lagrangian.} ultraviolet divergent contribution to
this Green's function can be generated in the classical statistical
approximation, then the CSA is not renormalizable.

Let us first consider this 4-point function at 1-loop. At this order,
the only possible contribution (up to trivial permutations of the
external legs) to the $\Gamma_{1112}$ function is
\setbox1\hbox to 3cm{\resizebox*{3cm}{!}{\includegraphics{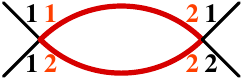}}\hfil}
\begin{equation}
-i\Gamma_{1112}^{\rm 1\ loop}=\raise -4mm\box1\; ,
\end{equation}
where the indices 1 and 2 indicate the various vertex
assignments. (The $-i$ prefactor is a convention, so that the function
$\Gamma$ can be viewed directly as a correction to the coupling
constant $g^2$.) Because it must contain a vertex of type $1112$, this
function is zero in the classical statistical
approximation\footnote{For the calculation of the full $\Gamma_{1112}$
  at one loop, beyond the classical statistical approximation, see the
  appendix \ref{app:1222-1112}.},
\begin{equation}
-i\big[\Gamma_{1112}\big]_{_{\rm CSA}}^{\rm 1\ loop}=0\; .
\end{equation}
Therefore, this 4-point function does not cause any renormalization
problem in the CSA at 1-loop. Similarly, the function $\Gamma_{1111}$
at one loop also requires the vertex $1112$, and is therefore zero in the
classical statistical approximation\footnote{At one loop, the
  functions $\Gamma_{1111}$ and $\Gamma_{2222}$ are also zero in the
  full theory.},
\begin{equation}
-i\big[\Gamma_{1111}\big]_{_{\rm CSA}}^{\rm 1\ loop}=0\; .
\end{equation}
For the function $\Gamma_{2222}$ at one loop, the only possibility is
the following, \setbox1\hbox to
3cm{\resizebox*{3cm}{!}{\includegraphics{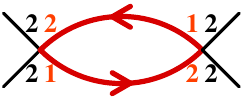}}\hfil}
\begin{equation}
-i\big[\Gamma_{2222}\big]_{_{\rm CSA}}^{\rm 1\ loop}=\raise -5mm\box1=0\; ,
\label{eq:2222-1loop}
\end{equation}
where we have represented with arrows the $12$ propagators, since they
are retarded propagators. This graphs is zero because it is made of a
sequence of retarded propagators forming a closed loop.

\subsubsection{Logarithmic divergence in $\Gamma_{1222}$}
At one loop, the function $\Gamma_{1222}$ is given by the graph of
eq.~(\ref{eq:1222-1loop}) (and several other permutations of the
indices).
\setbox1\hbox to 3cm{\resizebox*{3cm}{!}{\includegraphics{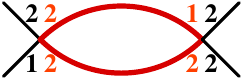}}\hfil}
\begin{equation}
-i\big[\Gamma_{1222}\big]_{_{\rm CSA}}^{\rm 1\ loop}=\raise -4mm\box1
\sim g^4 \log(\Lambda_{_{\rm UV}})\; .
\label{eq:1222-1loop}
\end{equation}
It is a straightforward calculation to check that this graph has a
logarithmic ultraviolet divergence, that can be removed by the
standard 1-loop renormalization of the coupling constant (this is
possible, since the interaction term $\phi_1\phi_2^3$ has been kept in
the Lagrangian when doing the classical statistical approximation).
The calculation of this 4-point function is detailed in the
appendix \ref{app:1222-1112}.

\subsubsection{Violation of Weinberg's theorem in $\Gamma_{1122}$}
Another interesting object to study is the 4-point function with
indices $1122$. There is no such bare vertex in the Lagrangian (both
for the unapproximated theory and for the classical statistical
approximation).  Since the full theory is renormalizable, this
function should not have ultraviolet divergences at 1-loop, since such
divergences would not be renormalizable. However, since the CSA
discards certain terms, it not obvious a priori that this conclusion
still holds. For the sake of definiteness, let us denote
$p_1,\cdots,p_4$ the external momenta of this function (defined to be
incoming into the graph, therefore $p_1+p_2+p_3+p_4=0$), and let us
assume that the two indices $1$ are attached to the legs $p_1,p_2$ and
the two indices $2$ are attached to the legs $p_3,p_4$. At one loop,
this 4-point function (in the full field theory) receives the
following contributions~:
\setbox1\hbox to 2.8cm{\resizebox*{2.8cm}{!}{\includegraphics{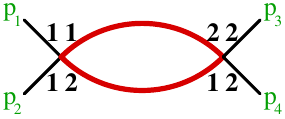}}}
\setbox2\hbox to 1.1cm{\resizebox*{1.1cm}{!}{\includegraphics{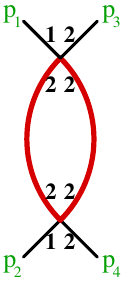}}}
\setbox3\hbox to 1.1cm{\resizebox*{1.1cm}{!}{\includegraphics{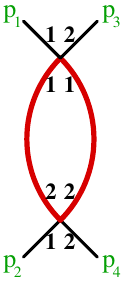}}}
\setbox4\hbox to 1.1cm{\resizebox*{1.1cm}{!}{\includegraphics{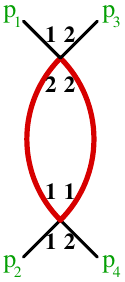}}}
\setbox5\hbox to 1.1cm{\resizebox*{1.1cm}{!}{\includegraphics{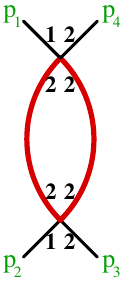}}}
\setbox6\hbox to 1.1cm{\resizebox*{1.1cm}{!}{\includegraphics{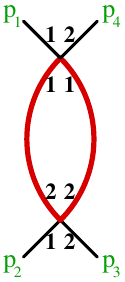}}}
\setbox7\hbox to 1.1cm{\resizebox*{1.1cm}{!}{\includegraphics{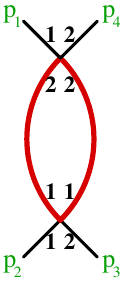}}}
\begin{eqnarray}
-i\Gamma_{1122}^{\rm 1\ loop}
&=
\underbrace{\raise -5mm\box1}_{\mbox{S channel}}
&+
\underbrace{
\raise -11mm\box2
+
\raise -11mm\box3
+
\raise -11mm\box4}_{\mbox{T channel}}
\nonumber\\
&&
+
\underbrace{
\raise -11mm\box5
+
\raise -11mm\box6
+
\raise -11mm\box7}_{\mbox{U channel}}\; .
\label{eq:1122-1loop}
\end{eqnarray}
Since the full theory is renormalizable, the sum of all these graphs
should be ultraviolet finite, because there is no $1122$ 4-field
operator in the bare Lagrangian.

It is however not obvious that the subset of these graphs that exist
in the classical statistical approximation is itself ultraviolet finite.
Among the T-channel and U-channel graphs, only the first of the three
graphs exist in the CSA, since all the other graphs contain the $1112$
bare vertex,
\setbox2\hbox to 1.1cm{\resizebox*{1.1cm}{!}{\includegraphics{G1122-T1}}}
\setbox5\hbox to 1.1cm{\resizebox*{1.1cm}{!}{\includegraphics{G1122-U1}}}
\begin{equation}
-i\big[\Gamma_{1122}\big]_{{\rm CSA}}^{\rm 1\ loop}
=
\raise -11mm\box2
+
\raise -11mm\box5\; .
\label{eq:1122-CSA}
\end{equation}
Some details of the calculation of these graphs are provided in the
appendix \ref{app:1122}. One obtains
\begin{equation}
-i\big[\Gamma_{1122}\big]_{{\rm CSA}}^{\rm 1\ loop}
=-\frac{g^4}{64\pi}\left[
{\rm sign}(T)+{\rm sign}(U)
+2\,\Lambda_{_{\rm UV}}
\left(
\frac{\theta(-T)}{|\p_1+\p_3|}
+
\frac{\theta(-U)}{|\p_1+\p_4|}
\right)
\right]\; ,
\label{eq:1122-1loop-1}
\end{equation}
where we denote
\begin{equation}
T\equiv (p_1+p_3)^2\quad,\quad U\equiv (p_1+p_4)^2\; ,
\end{equation}
and where $\Lambda_{_{\rm UV}}$ is an ultraviolet cutoff introduced to
regularize the integral over the 3-momentum running in the
loop. As one sees, these graphs have a {\sl linear} ultraviolet
divergence, despite having a superficial degree of divergence equal to
zero. This property violates Weinberg's theorem since, if it were
applicable here, it would imply at most a logarithmic divergence with
a coefficient independent of the external momenta. One can attribute
this violation to the analytic structure of the integrand\footnote{For
  the graphs in eq.~(\ref{eq:1122-CSA}), the integrand is of the form
  $\delta(K^2)\delta((P+K)^2)$ where $P\equiv p_1+p_3$ or $P\equiv
  p_1+p_4$. Using the first delta function, the argument of the second
  one is $(P+K)^2=2P\cdot K+P^2$, which is only of degree $1$ in the
  loop momentum. Therefore, the second propagator contributes only
  $-1$ to the actual degree of divergence of the graph, contrary to
  the $-2$ assumed based on dimensionality when computing the
  superficial degree of divergence. This discrepancy is also related to the
  impossibility to perform a Wick's rotation when the integrand is
  expressed in terms of delta functions or retarded/advanced
  propagators.}: unlike in ordinary Feynman perturbation theory, we
cannot perform a Wick rotation to convert the integral to an integral
over an Euclidean momentum, which is an important step in the proof of
Weinberg's theorem. 

Since it occurs in the operator $\phi_1^2\phi_2^2$, that does not
appear in the CSA Lagrangian, this linear divergence provides
incontrovertible proof of the fact that {\sl the classical statistical
  approximation is not renormalizable}. Moreover, this conclusion is
independent of the value of the coupling constant. The only thing one
gains at smaller coupling is that the irreducible cutoff dependence
caused by these terms is weaker.

It should also be noted that this linear divergence is a purely
imaginary contribution to the function $\Gamma_{1122}$ (this can be
understood from the structure of the integrand, that was made of two
delta functions, which is reminiscent of the calculation of the
imaginary part of a Green's function via Cutkosky's cutting rules
\cite{Cutko1,t'HooV1}).

In the appendix \ref{app:1122}, we also calculate the graphs of
eq.~(\ref{eq:1122-1loop}) that do not contribute to the CSA, and we
find
\setbox3\hbox to 1.1cm{\resizebox*{1.1cm}{!}{\includegraphics{G1122-T2}}}
\setbox4\hbox to 1.1cm{\resizebox*{1.1cm}{!}{\includegraphics{G1122-T3}}}
\setbox6\hbox to 1.1cm{\resizebox*{1.1cm}{!}{\includegraphics{G1122-U2}}}
\setbox7\hbox to 1.1cm{\resizebox*{1.1cm}{!}{\includegraphics{G1122-U3}}}
\begin{eqnarray}
\raise -11mm\box3
+
\raise -11mm\box4
+
\raise -11mm\box6
+
\raise -11mm\box7
&=&
-\frac{g^4}{32\pi}\Big[
1
\nonumber\\
&&
\!\!\!\!-\Lambda_{_{\rm UV}}
\left(
\frac{\theta(-T)}{|\p_1+\p_3|}
+
\frac{\theta(-U)}{|\p_1+\p_4|}
\right)
\Big]\; .\nonumber\\
&&
\label{eq:1122-1loop-2}
\end{eqnarray}
(Note that the S-channel graph is in fact zero, because it is made of
a sequence of retarded propagators in a closed loop.) By adding
eqs.~(\ref{eq:1122-1loop-1}) and (\ref{eq:1122-1loop-2}), we obtain
the 1-loop result in the unapproximated theory,
\begin{equation}
-i\Gamma_{1122}^{\rm 1\ loop}=-\frac{g^4}{32\pi}\left[\theta(T)+\theta(U)\right]\; ,
\end{equation}
which is ultraviolet finite, in agreement with the renormalizability
of the full theory.

\subsection{Two point functions at two loops}
Let us also mention two problematic 2-point functions at two loops. We
just quote the results here (the derivation will be given in
\cite{EpelbGTW1}), for an on-shell momentum $P$ ($P^2=0,p_0>0$)~:
\setbox1\hbox to
3.5cm{\resizebox*{3.5cm}{!}{\includegraphics{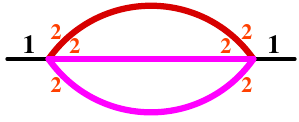}}}
\begin{eqnarray}
-i\big[\Sigma_{11}(P)\big]_{_{\rm CSA}}^{\rm 2\ loop}
&=&
\raise -6.5mm\box1
\nonumber\\
&=&
-\frac{g^4}{1024\pi^3}\;\left(\Lambda_{_{\rm UV}}^2-\frac{2}{3}p^2\right)\; ,
\label{eq:S11-2loop}
\end{eqnarray}
\setbox1\hbox to 3.5cm{\resizebox*{3.5cm}{!}{\includegraphics{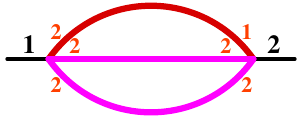}}}
\begin{eqnarray}
{\rm Im}\,\big[\Sigma_{12}(P)\big]_{_{\rm CSA}}^{\rm 2\ loop}
&=&
\raise -6.5mm\box1
\nonumber\\
&=&
-\frac{g^4}{1024\pi^3}
\;\left(\Lambda_{_{\rm UV}}^2-\frac{2}{3}p^2\right)\; .
\label{eq:S12-2loop}
\end{eqnarray}
An ultraviolet divergence in $\Sigma_{11}$ is non-renormalizable,
since there is no $\phi_1^2$ operator in the Lagrangian. Similarly,
the divergence at 2-loops in ${\rm Im}\,\Sigma_{12}$ is also
non-renormalizable, because it would require an imaginary counterterm,
that would break the Hermiticity of the Lagrangian.

\section{Consequences on physical observables}
\label{sec:int}
\subsection{Order of magnitude of the pathological terms}
So far, we have exhibited a 4-point function at 1-loop that has an
ultraviolet divergence in the CSA but not if computed in full, and
that cannot be renormalized in the CSA because it would require a
counterterm for an operator that does not exist in the Lagrangian.  

In practice, this 1-loop function enters as a subdiagram in the loop
expansion of observable quantities, making them unrenormalizable. In
order to assess the damage, it is important to know the lowest order
at which this occurs. Let us consider in this discussion two
quantities that have been commonly computed with the classical
statistical method: the expectation value of the energy-momentum
tensor, and the occupation number, which can be extracted from the
$G_{22}$ propagator.
\setbox1\hbox to 4cm{\resizebox*{4cm}{!}{\includegraphics{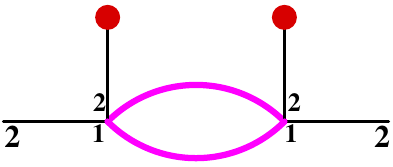}}}
\setbox2\hbox to 3cm{\resizebox*{3cm}{!}{\includegraphics{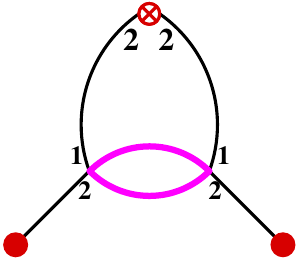}}}

For the $22$ component of the propagator, the first occurrence of the
$1122$ 4-point function as a subgraph is in the following 1-loop
contribution~:
\begin{equation}
\left[G_{22}\right]{}_{\rm CSA}^{^{\rm NNLO}} = \raise -4mm\box1\; ,
\label{eq:G22-NNLO}
\end{equation}
where the problematic subdiagram has been highlighted (the propagators
in purple are of type $G_{22}$). The fact that this problematic
subgraphs occurs only at 1-loop and beyond is related to the fact that
classical statistical approximation is equivalent to the full theory
up to (including) NLO. The first differences appear at NNLO, which
means 1-loop for the $G_{22}$ function. In a situation where the
typical physical scale is denoted $Q$, the subdiagram is of order $g^4
\Lambda_{_{\rm UV}}/Q$, and the external field attached to the graph
is of order $\Phi_2\sim Q/g$ (we assume a system dominated by strong
fields, as in applications to heavy ion collisions). This 1-loop
contribution to $G_{22}$ is of order $g^2\Lambda_{_{\rm UV}} Q$, to be
compared to $Q^2/g^2$ at leading order. Therefore, the relative
suppression of this non-renormalizable contribution is by a factor
\begin{equation}
g^4\frac{\Lambda_{_{\rm UV}}}{Q}\; .
\label{eq:rel}
\end{equation}

The same conclusion holds in the case of the energy-momentum tensor,
for which the $1122$ 4-point function enters also at NNLO (in this
case, this means two loops), in the following diagram~:
\begin{equation}
\left[T^{\mu\nu}\right]{}^{^{\rm NNLO}}_{{\rm CSA}} = \raise -13mm\box2\; .
\label{eq:T-NNLO}
\end{equation}
(The cross denotes the insertion of the $T^{\mu\nu}$ operator.) The
order of magnitude of this graph is $g^2\Lambda_{_{\rm UV}}Q^3$, while the
leading order contribution to the energy-momentum tensor is of order
$Q^4 / g^2$. Therefore, the relative suppression is the same as in
eq.~(\ref{eq:rel}).  

All these examples suggest that a minimum requirement is that the
ultraviolet cutoff should satisfy
\begin{equation}
\Lambda_{_{\rm UV}}\ll \frac{Q}{g^4}\; ,
\label{eq:cond0}
\end{equation}
for the above contributions to give only a small contamination to
their respective observables in a classical statistical computation
with cutoff $\Lambda_{_{\rm UV}}$. However, one could be a bit more
ambitious and request that this computation be also accurate at
NLO. For this, we should set the cutoff so that the above diagrams are
small corrections compared to the NLO contributions. This is achieved
if the highlighted 4-point function in these graphs is small compared
to the tree-level 4-point function, i.e. $g^2$. This more stringent
condition reads
\begin{equation}
\frac{g^4}{16\pi}\frac{\Lambda_{_{\rm UV}}}{Q}\ll g^2\quad,\quad\mbox{i.e.}\;\; \Lambda_{_{\rm UV}}\ll \frac{16\pi Q}{g^2}\; ,
\label{eq:cond1}
\end{equation}
where we have reintroduced the factors $2$ and $\pi$ from
eq.~(\ref{eq:1122-1loop-1}), because in practical situations they are
numerically important. One can see that this inequality is easy to
satisfy at weak coupling $g^2\ll 1$, and presumably only marginally satisfied at
larger couplings $g\approx 1$.

\subsection{Ultraviolet contamination at asymptotic times}
The condition of eq.~(\ref{eq:cond1}) ensures that the pathological
NNLO contributions are much smaller than the NLO corrections (the
latter are correctly given by the classical statistical
approximation). Another important aspect of this discussion is
whether, by ensuring that the inequality (\ref{eq:cond1}) is
satisfied, one is guaranteed that the contamination by the
pathological terms remains small at all times. It is easy to convince
oneself that this is not the case. In Ref.~\cite{EpelbGTW1}, we argue
that these pathological terms, if not removed, induce corrections that
become comparable to the physical result after a time that varies as
$Qt_*\sim 2048\pi^3 g^{-4} (Q/\Lambda_{_{\rm UV}})^2$. Effectively,
these ultraviolet divergent terms act as spurious scatterings with a
rate proportional to $g^4 \Lambda_{_{\rm UV}}^2/Q$.

Moreover, the state reached by the system when $t\to+\infty$ is
controlled solely by conservation laws and by a few quantities that
characterize the initial condition, in addition to the ultraviolet
cutoff. For instance, if the only conserved quantities are energy and
momentum, then the asymptotic state depends only on the total energy
in the system. If in addition the particle number was conserved, then
the asymptotic state would also depend on the number of particles in
the system.

In particular, the value of the coupling constant does not play a role
in determining which state is reached at asymptotic times; it only
controls how quickly the system approaches the asymptotic state. This
means that, even if $g^2$ is small so that the inequality
(\ref{eq:cond1}) is satisfied, the CSA may evolve the system towards
an asymptotic state that differs significantly from the true
asymptotic state, regardless of how small $g^2$ is. Therefore, the
strong dependence of the asymptotic state on the ultraviolet cutoff
observed in the figure 10 of Ref.~\cite{BergeBSV3} is not specific to
a ``large'' coupling $g^2=1$. Exactly the same cutoff dependence would
be observed at smaller couplings, but the system would need to evolve
for a longer time in order to reach it.

\subsection{Could it be fixed?}
An important issue is whether one could somehow alter the classical
statistical approximation in order to remove the linear divergence
that appears in the 1-loop 4-point function $\Gamma_{1122}$. As a
support of these considerations, let us consider the NNLO correction
to the function $G_{22}$. Eq.~(\ref{eq:G22-NNLO}) displays the unique
contribution in the classical statistical approximation. However, in
the full theory there are two other possible arrangements of the
internal $1/2$ inside the $1122$ subdiagram. This topology with the
complete $1122$ subdiagram reads~:
\setbox1\hbox to 11cm{\resizebox*{11cm}{!}{\includegraphics{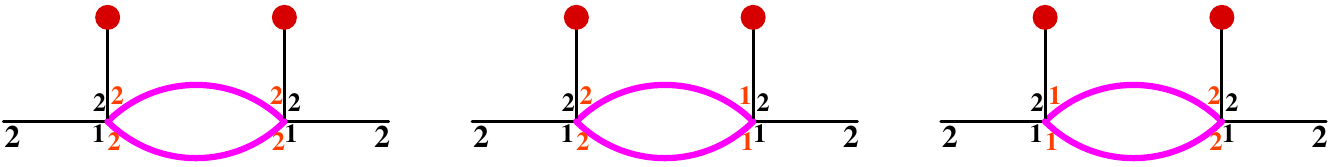}}}
\begin{equation}
\raise -3.5mm\box1\; .
\label{eq:G22-NNLO-full}
\end{equation}
The CSA result contains only the first graph, and as a consequence it
has a linear ultraviolet divergence, while the sum of the three graphs
is finite. So the question is: could one reintroduce in the CSA the
divergent part of the 2nd and 3rd graphs, in order to compensate the
divergence of the first graph?

\begin{figure}[htbp]
\begin{center}
\resizebox*{!}{2.8cm}{\includegraphics{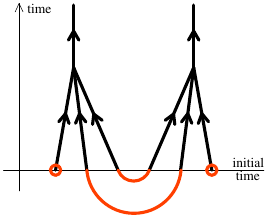}}
\end{center}
\caption{\label{fig:G22-CSA}Space-time representation of
  eq.~(\ref{eq:G22-NNLO}). The propagators with an arrow are retarded
  propagators. The orange circles represent the mean value of the
  initial field. The orange lines represent the link coming from the
  Gaussian fluctuations of the initial field.}
\end{figure}

In order to better visualize what it would take to do this, let us
modify the way the graphs are represented, so that they reflect
the space-time evolution of the system and the {\sl modus operandi} of
practical implementations of the CSA. The modified representation for
the first term in eq.~(\ref{eq:G22-NNLO-full}) is shown in the figure
\ref{fig:G22-CSA}.
In this representation, the lines with arrows are retarded propagators
(the time flows in the direction of the arrow). The solution of the
classical equation of motion is a sum of trees made of retarded
propagators, where the ``leaves'' of the tree are anchored to the
initial surface. In the diagram shown in the figure \ref{fig:G22-CSA},
there are two such trees, both containing one instance of the quartic
interaction term. In order to complete the calculation in the CSA, one
performs a Gaussian average over the initial value of the classical
field. Diagrammatically, this average amounts to attaching the leaves
of the tree to 1-point objects representing the average value of the
initial field, or to connecting them pairwise with the 2-point
function that describes the variance of the initial Gaussian
distribution.  

It is crucial to note that the trees that appear in the solution of
the classical equation of motion are ``oriented''~: three retarded
propagators can merge at a point, from which a new retarded propagator
starts.  Let us call this a $3\to 1$ vertex (when read in the
direction of increasing time). These trees do not contain any
$2\to 2$ or $1\to 3$ vertices. Their absence is intimately related to
the absence of the $1122$ and $1112$ vertices in the Lagrangian in the
classical statistical approximation.

\begin{figure}[htbp]
\begin{center}
\resizebox*{!}{3cm}{\includegraphics{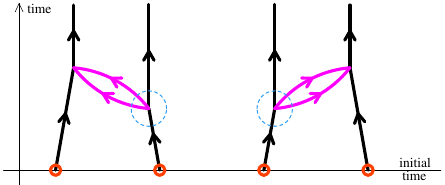}}
\end{center}
\caption{\label{fig:G22-NCSA}Space-time representation of the NNLO
  contributions to $G_{22}$ that are not included in the classical
  statistical approximation. The dotted circles outline the $1112$
  vertices, that are missing in the CSA.}
\end{figure}

\begin{figure}[htbp]
\begin{center}
\resizebox*{!}{3cm}{\includegraphics{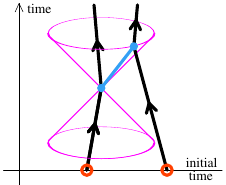}}
\end{center}
\caption{\label{fig:G22-CT}Space-time representation of the divergent
  part of the graphs of figure \ref{fig:G22-NCSA}. As explained in the
  text, these divergent terms are non-local in space-time, with
  support on the light-cone.}
\end{figure}

In the figure \ref{fig:G22-NCSA}, we now show the same representation
for the 2nd and 3rd contributions of eq.~(\ref{eq:G22-NNLO-full}).
Firstly, we see that these graphs contain a $1\to 3$ vertex
(surrounded by a dotted circle in the figure), in agreement with the
fact that they do not appear in the CSA.
There is no way to generate the loop contained in this graphs via the
average over the initial conditions, because this loop corresponds to
quantum fluctuations that happen later on in the time
evolution. By Fourier transforming the divergent part of
these diagrams in eq.~(\ref{eq:1122-1loop-2}), we can readily see that
it is proportional to
\begin{equation}
\frac{1}{|\x-\y|}\,\delta((x^0-y^0)^2-(\x-\y)^2)
\end{equation}
in coordinate space. Thus, the divergent part of these loops is
non-local, with support on the light-cone, as illustrated in the
figure \ref{fig:G22-CT}.

\begin{figure}[htbp]
\begin{center}
\resizebox*{!}{4cm}{\includegraphics{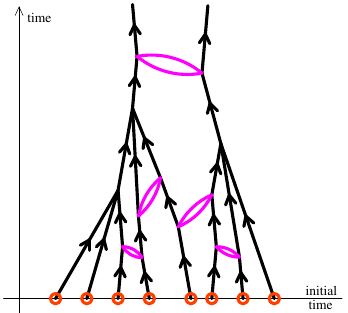}}
\end{center}
\caption{\label{fig:G22-multi}Contribution to $G_{22}$ with many
  $\Gamma_{1122}$ subgraphs.}
\end{figure}

There can also be arbitrarily many occurrences of these divergent
subgraphs in the calculation of an observable in the classical
statistical method, as illustrated in the figure \ref{fig:G22-multi}
in the case of $G_{22}$.
This implies that these divergences cannot be removed by an overall
subtraction, and that one must instead modify the Lagrangian. One
could formally subtract them by adding to the action of the theory a
non-local counterterm\footnote{Of course, there was no
  $\phi_1^2\phi_2^2$ term in the original bare action. On the other
  hand, we know that in the full theory, there should not be an
  intrinsic ultraviolet divergence in the $1122$ function. One should
  view this counterterm as a way of reintroducing some of the terms
  that are beyond the classical statistical approximation, in order to
  restore the finiteness of the $1122$ function.} of the form
\begin{equation}
\Delta{\cal S}\equiv
-\frac{i}{2}\int\rmd^4x\,\rmd^4y\; \left[\phi_1(x)\phi_2(x)\right]\,
v(x,y)\,
\left[\phi_1(y)\phi_2(y)\right]\; ,
\end{equation}
where 
\begin{equation}
v(x,y)\equiv \frac{g^4}{64\pi^3}\frac{\Lambda_{_{\rm UV}}}{|\x-\y|}\,
\delta((x^0-y^0)^2-(\x-\y)^2)
\end{equation}
is tuned precisely to cancel the linear divergence in the
$\Gamma_{1122}$ function.  In order to deal with such a term, the
simplest is to perform a {\sl Hubbard-Stratonovich} transformation
\cite{Hubba1,Strat1}, by introducing an auxiliary field $\zeta(x)$
via the following identity
\begin{equation}
e^{i\Delta{\cal S}}
=
\int[{\rm D}\zeta]\;
e^{\frac{1}{2}\int_{x,y}\zeta(x)v^{-1}(x,y)\zeta(y)}
\;
e^{i\int_x \zeta(x)\,\phi_1(x)\phi_2(x)}\; .
\end{equation}
The advantage of this transformation is that we have transformed a
non-local four-field interaction term into a local interaction with a
random Gaussian auxiliary field.

\begin{figure}[htbp]
\begin{center}
\resizebox*{!}{4cm}{\includegraphics{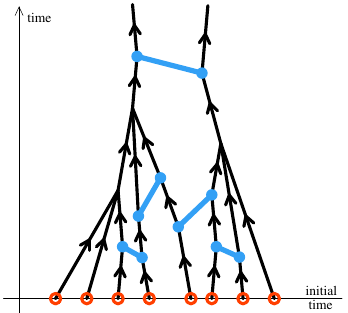}}
\end{center}
\caption{\label{fig:G22-multi-CT}Effect of the noise term on the
  topology shown in the figure \ref{fig:G22-multi}.}
\end{figure}

The rest of the derivation of the classical statistical method remains
the same: the field $\phi_1$ appears as a Lagrange multiplier for a
classical equation of motion for the field $\varphi$, but now we get an
extra, stochastic, term in this equation~:
\begin{equation}
  (\square+m^2)\varphi+\frac{g^2}{6}\varphi^3 + i\xi \varphi=j\; .
  \label{eq:CL}
\end{equation}
Note that we have introduced $\zeta\equiv i\xi$ in order to have a
positive definite variance for the new variable $\xi$. From the
above derivation, this noise must be Gaussian distributed, with a mean
and variance given by the following formulas,
\begin{eqnarray}
\big<\xi(x)\big>&=&0\nonumber\\
\big<\xi(x)\xi(y)\big>&=& 
\frac{g^4}{64\pi^3}
\frac{\Lambda_{_{\rm UV}}}{|\x-\y|}\,\delta((x^0-y^0)^2-(\x-\y)^2)
\; .
\label{eq:noise}
\end{eqnarray}
By construction, the noise term in eq.~(\ref{eq:CL}), once averaged
with eq.~(\ref{eq:noise}), will insert a non-local counterterm in
every place where the $\Gamma_{1122}$ function can appear. For
instance, when applied to the calculation of $G_{22}$, the
contribution shown in the figure \ref{fig:G22-multi} will be
accompanied by the term shown in the figure \ref{fig:G22-multi-CT}.

Although the noise in eq.~(\ref{eq:noise}) has non-local space-time
correlations, it is easy to generate it in momentum space, where it
becomes diagonal. The main practical difficulty however comes from the
non-locality in time\footnote{This non-locality appears to be a
  reminiscence of the memory effects that exist in the full quantum
  field theory, but are discarded in the classical statistical
  approximation. Note that the 2PI resummation scheme also has such
  terms.}: one would need to generate and store the whole
spatio-temporal dependence for each configuration of the noise term
prior to solving the modified classical equation of motion.

The noise term introduced in eq.~(\ref{eq:CL}) is purely imaginary,
and it turns the classical field $\phi$ into a complex valued
quantity. However, since $\xi$ is Gaussian distributed with a zero
mean, any Hermitean observable constructed from $\phi$ via an average
over $\xi$ will be real valued\footnote{The average over $\xi$ will
  only retain terms that are even in $\xi$, and the factors $i$ will
  cancel.}. Eq.~(\ref{eq:CL}) is therefore a complex Langevin
equation, and may be subject to the problems sometimes encountered
with this kind of equations (lack of convergence, or convergence to
the incorrect solution). At the moment, it is an open question whether
eq.~(\ref{eq:CL}) really offers a practical way of removing the linear
ultraviolet divergences from the classical statistical approximation.

\section{Conclusions and outlook}
\label{sec:concl}
In this work, we have investigated the ultraviolet behavior of the
classical statistical approximation. This has been done by using
perturbation theory in the retarded-advanced basis, where this
approximation has a very simple expression, and in calculating all the
one-loop subdiagrams that can possibly be generated with these
diagrammatic rules.

The main conclusion of this study is that the classical approximation
leads to a 1-loop 4-point function that diverges linearly in the
ultraviolet cutoff. More specifically, the problem lies in the
function $\Gamma_{1122}$, where the $1,2$ indices refer to the
retarded-advanced basis. In the unapproximated theory, this function
is ultraviolet finite, but it violates Weinberg's theorem in the
classical statistical approximation, because it has an ultraviolet
divergence with a coefficient which is non-polynomial in the external
momenta. Moreover, it is non-renormalizable because it corresponds to
an operator that does not even appear in the Lagrangian one started
from.

The mere existence of these divergent terms implies that the classical
statistical approximation is not renormalizable, no matter what the
value of the coupling constant is. 

We have estimated that the contamination of the results by these
non-renormalizable terms is of relative order $g^4\Lambda_{_{\rm
    UV}}/Q$, where $Q$ is the typical physical momentum scale of the
problem under consideration.  Based on this, the general rule is that
the coupling should not be too large, and the cutoff should remain
close enough to the physical scales. 

In this paper, we have also proposed that this one-loop spurious
(because it does not exist in the full theory) divergence may be
subtracted by adding a multiplicative Gaussian noise term to the
classical equation of motion. This noise term can be tuned in order to
reintroduce some of the terms of the full theory that had been lost
when doing the classical approximation. In order to subtract the
appropriate quantity, this noise must be purely imaginary, with a 
2-point correlation given by the Fourier transform of the divergent
term.  Unfortunately, this correlation is non-local in time, which
makes the implementation of this correction quite complicated. Whether
this can be done in practice remains an open question at this point.

Moreover, the ultraviolet contamination due to these
non-renormalizable terms is cumulative over time, and will eventually
dominate the dynamics of the system no matter how small
$g^4\Lambda_{_{\rm UV}}/Q$ is. An extensive discussion of this
asymptotic ultraviolet sensitivity, and of the time evolution that
leads to the asymptotic state, will be provided in a forthcoming work
\cite{EpelbGTW1}.

\section*{Acknowledgements}
We would like to thank L.~McLerran for useful discussions about
this work and closely related issues.  This work is supported by the
Agence Nationale de la Recherche project 11-BS04-015-01.

\appendix

\section{$\Gamma_{1222}$ and $\Gamma_{1112}$ at one loop}
\label{app:1222-1112}
At one loop, the 4-point functions $\Gamma_{1222}$ and $\Gamma_{1112}$ 
are given by the following sets of graphs~: 
\setbox1\hbox to 2.8cm{\resizebox*{2.8cm}{!}{\includegraphics{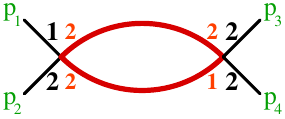}}}
\setbox2\hbox to 1.1cm{\resizebox*{1.1cm}{!}{\includegraphics{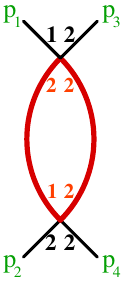}}}
\setbox3\hbox to 1.1cm{\resizebox*{1.1cm}{!}{\includegraphics{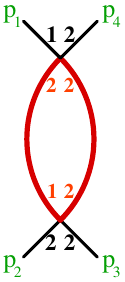}}}
\setbox4\hbox to 2.8cm{\resizebox*{2.8cm}{!}{\includegraphics{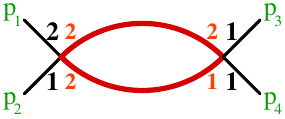}}}
\setbox5\hbox to 1.1cm{\resizebox*{1.1cm}{!}{\includegraphics{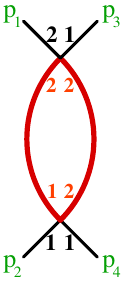}}}
\setbox6\hbox to 1.1cm{\resizebox*{1.1cm}{!}{\includegraphics{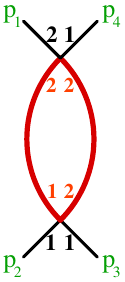}}}
\begin{eqnarray}
-i\Gamma_{1222}^{\rm 1\ loop}(p_1\cdots p_4)
&=&
\raise -5mm\box1
+
\raise -11mm\box2
+
\raise -11mm\box3\nonumber\\
&=&
-i\left[I(p_3+p_4)+I(p_2+p_3)+I(p_2+p_4)\right]\; ,
\end{eqnarray}
\begin{eqnarray}
-i\Gamma_{2111}^{\rm 1\ loop}(p_1\cdots p_4)
&=&
\raise -5mm\box4
+
\raise -11mm\box5
+
\raise -11mm\box6\nonumber\\
&=&
-\frac{i}{4}\left[I(p_3+p_4)+I(p_2+p_3)+I(p_2+p_4)\right]\; .
\end{eqnarray}
Note that all the 1-loop contributions to $\Gamma_{1112}$ are zero if
we exclude the $1112$ vertex, as done in the classical statistical
approximation. The fact that $\Gamma_{1222}$ and $\Gamma_{2111}$
differ only by a factor $1/4$ is a consequence of their common origin
in the Schwinger-Keldysh formalism, where the $++++$ and $----$ vertex
functions are mutual complex conjugates.

They are all expressible in terms of a single loop integral,
\begin{equation}
I(P)
\equiv
-ig^4
\int\frac{\rmd^4K}{(2\pi)^4}\;G_{22}(K)\,G_{12}(P+K)\; .
\end{equation}
Since $G_{22}(K)$ is a delta function $\delta(K^2)$, we use it in
order to perform the integration over the energy $k^0$. Then, the
integration over $\cos(\theta)$, where $\theta$ is the angle between
the 3-vectors $\p$ and $\k$, is elementary but requires that one
studies carefully whether $P^2+2P\cdot K$ can vanish in the
integration range. If this quantity can vanish, the integral will also
have an imaginary part.  This leads to the following expression for
$I(P)$~:
\begin{eqnarray}
I(P)
&=&
\frac{g^4}{32\pi^2}
\smash{\Bigg\{
\frac{1}{p}\int\limits_0^{\Lambda_{_{\rm UV}}}}
\rmd k\;\sum_{\epsilon=\pm 1}
\log\left|\frac{p+\epsilon p^0+P^2/2k}{p-\epsilon p^0-P^2/2k}\right|
\nonumber\\
&&\qquad\qquad\qquad
+i\pi\left(\theta(P^2){\rm sign}(p^0)-\theta(-P^2)\frac{p^0}{p}\right)
\smash{\Bigg\}}\nonumber\\
&=&
\frac{g^4}{32\pi^2}
\smash{\Bigg\{}
\ln\left|\frac{P^2}{4\Lambda_{_{\rm UV}}^2}\right|
+
\frac{p^0}{p}\log\left|\frac{p^0+p}{p^0-p}\right|
-2
\nonumber\\
&&\qquad\qquad\qquad
+i\pi\left(\theta(P^2){\rm sign}(p^0)-\theta(-P^2)\frac{p^0}{p}\right)
\smash{\Bigg\}}\; .
\end{eqnarray}
Therefore, the real part is logarithmically ultraviolet divergent,
while the imaginary part is finite.

\section{$\Gamma_{1122}$ at one loop}
\label{app:1122}
In this appendix, we perform the calculation of some of the graphs
contributing to $\Gamma_{1122}^{\rm 1\ loop}$. The list of all the
relevant graphs is given in eq.~(\ref{eq:1122-1loop}). The unique
S-channel graph is zero, because it has a closed loop of retarded
propagators. Since the T- and U-channel graphs are identical, up to
the permutation $p_3\leftrightarrow p_4$, we will calculate only the
T-channel graphs here.

\subsection{Classical statistical approximation at one-loop}

Let us consider first the graph the contributes in the classical
statistical approximation~:
\setbox2\hbox to 1.1cm{\resizebox*{1.1cm}{!}{\includegraphics{G1122-T1}}}
\begin{equation}
\raise -11mm\box2
=
-\frac{g^4}{2} \int \frac{\rmd^4K}{(2\pi)^4}\;
\pi\delta(K^2)\pi\delta((P-K)^2)\; ,
\label{eq:1122-cut1}
\end{equation}
where we denote $P\equiv p_1+p_3$. We first use the $\delta(K^2)$ in order to perform
the integral over $k^0$, which gives
\setbox2\hbox to 1.1cm{\resizebox*{1.1cm}{!}{\includegraphics{G1122-T1}}}
\begin{equation}
  \raise -11mm\box2
  =
  -\frac{g^4}{32\pi} \int\limits_0^{\Lambda_{_{\rm UV}}} k \rmd{k} \int\limits_{-1}^{+1}\rmd(\cos\theta)\;
  \sum_{\epsilon=\pm}\delta(P^2-2\epsilon k p^0+2pk\cos\theta)\; ,
\label{eq:1122-cut}
\end{equation}
Anticipating the fact that the integral over $k=|\k|$ is
ultraviolet divergent, we have introduced an upper cutoff on this
integral. The second step is to use the remaining delta function in
order to integrate over $\cos\theta$. This requires some careful
analysis, in order to determine whether there is a valid solution,
i.e. one for which $-1\le\cos\theta\le +1$. This depends on the sign
$\epsilon$, on the sign of $P^2$ and on the value of $k$. The results
are summarized here~:
\begin{itemize}
\item If $P^2>0$, there is a valid solution if and only if~:
  \begin{equation*}
    \epsilon p^0>0
    \quad\mbox{and}\quad
    \frac{|p^0|-p}{2}\le k\le \frac{|p^0|+p}{2}\; ,
  \end{equation*}
\item If $P^2<0$~:
  \begin{itemize}
  \item There is no solution if $k<\frac{p-|p^0|}{2}$
  \item If $\epsilon p^0<0$, there is a solution if $\frac{p-|p^0|}{2}\le k$
  \item If $\epsilon p^0>0$, there is a solution if $\frac{p+|p^0|}{2}\le k$
  \end{itemize}
\end{itemize}
Using this, the integration over $\cos\theta$ leads to a piecewise
constant integrand for the remaining integral. We eventually obtain
\setbox2\hbox to 1.1cm{\resizebox*{1.1cm}{!}{\includegraphics{G1122-T1}}}
\begin{equation}
  \raise -11mm\box2
  =-\frac{g^4}{64\pi}\left[
{\rm sign}(P^2)
+2\,\Lambda_{_{\rm UV}}
\frac{\theta(-P^2)}{|\P|}
\right]\; .
\label{eq:1122-cut2}
\end{equation}

\subsection{One-loop graphs beyond the CSA}

Let us now focus on the 2nd and 3rd T-channel graphs,
\setbox3\hbox to 1.1cm{\resizebox*{1.1cm}{!}{\includegraphics{G1122-T2}}}
\setbox4\hbox to 1.1cm{\resizebox*{1.1cm}{!}{\includegraphics{G1122-T3}}}
\begin{equation}
\raise -11mm\box3
+
\raise -11mm\box4
=
-\frac{g^4}{8}\int\frac{\rmd^4K}{(2\pi)^4}\;\left[G_{_R}(K)G_{_R}(P-K)+G_{_A}(K)G_{_A}(P-K)\right]\; .
\end{equation}
Changing $K^\mu\to -K^\mu$ in the second term, we see that this term
is the same as the first one with the change $P^\mu\to
-P^\mu$. Therefore, we need only to calculate the first term, multiply
by two its $P$-even part and discard its $P$-odd part. The first step
is to perform the $k^0$ integral in the complex plane. By closing the
integration contour in the lowest half-plane, we pick the two poles of
$G_{_R}(K)$, and we get
\setbox3\hbox to 1.1cm{\resizebox*{1.1cm}{!}{\includegraphics{G1122-T2}}}
\begin{equation}
  \raise -11mm\box3
  =
  -\frac{g^4}{64\pi^2}
  \int\limits_0^{\Lambda_{_{\rm UV}}} k \rmd{k} \int\limits_{-1}^{+1}\rmd(\cos\theta)\;
  \sum_{\epsilon=\pm}\epsilon
  \left[G_{_R}(P-K)\right]_{k^0=\epsilon k}\; .
\end{equation}
At this point, one can decompose the retarded propagator into a
principal value term and a delta function. One can check that the
principal value gives only terms that are $P$-odd, that we can thus
drop. Keeping only the delta function leads to
\setbox3\hbox to 1.1cm{\resizebox*{1.1cm}{!}{\includegraphics{G1122-T2}}}
\begin{equation}
  \raise -11mm\box3
  =
  -\frac{g^4}{64\pi}
  \int\limits_0^{\Lambda_{_{\rm UV}}} k \rmd{k} \int\limits_{-1}^{+1}\rmd(\cos\theta)\;
  \sum_{\epsilon=\pm}
  {\rm sign}(\epsilon p^0-k)\delta(P^2-2\epsilon k p^0+2pk\cos\theta)\; .
\end{equation}
We proceed by using the delta function to perform the integral over
$\cos\theta$. The conditions for having a valid solution are the same
as before. Finally, we obtain
\setbox3\hbox to 1.1cm{\resizebox*{1.1cm}{!}{\includegraphics{G1122-T2}}}
\setbox4\hbox to 1.1cm{\resizebox*{1.1cm}{!}{\includegraphics{G1122-T3}}}
\begin{equation}
\raise -11mm\box3
+
\raise -11mm\box4
=
-\frac{g^4}{32\pi}\left[
\frac{1}{2}
-\Lambda_{_{\rm UV}}
\frac{\theta(-P^2)}{|\P|}
\right]
\end{equation}

\subsection{CSA result with symmetric regularization}
The ultraviolet regularization introduced in eq.~(\ref{eq:1122-cut})
is not entirely satisfactory, because it breaks the symmetry between
the two internal lines of the graph, by placing a cutoff
$\Lambda_{_{\rm UV}}$ on the 3-momentum $|\k|$, while the 3-momentum
$|\p-\k|$ in the other internal line is not constrained. 

It is perfectly fine to do so.  However, some important identities
obeyed by 2-loop self-energies, in which the 1-loop $\Gamma_{1122}$
function appears as a subgraph, rely on the symmetry between the
internal lines of the graph. It is therefore important that the
regularization scheme employed in intermediate steps of the
calculation does not break this symmetry, and in particular that the
subgraph itself respects this symmetry.  This issue will be discussed
at length in a forthcoming work, Ref.~\cite{EpelbGTW1}, but for later
reference we present here the formulas for the function
$\Gamma_{1122}$ at 1-loop with a symmetric regularization. This just
amounts to replacing eq.~(\ref{eq:1122-cut}) by \setbox2\hbox to
1.1cm{\resizebox*{1.1cm}{!}{\includegraphics{G1122-T1}}}
\begin{eqnarray}
  &&\nonumber\\
  &&\nonumber\\
  \smash{\raise -11mm\box2}
  &=&
  -\frac{g^4}{32\pi} \int k \rmd{k} \smash{\int\limits_{-1}^{+1}}\rmd(\cos\theta)\;
  \sum_{\epsilon=\pm}\delta(P^2-2\epsilon k p^0+2pk\cos\theta)\nonumber\\
  &&\qquad\qquad\qquad\qquad\times\,
  \theta(\Lambda_{_{\rm UV}}-|\k|)\,
  \theta(\Lambda_{_{\rm UV}}-|\p-\k|)\; ,
\label{eq:1122-cut-sym}
\end{eqnarray}
with $|\p-\k|=(p^2+k^2-2pk\cos\theta)^{1/2}$. The new constraint on
$|\p-\k|$ slightly complicates the discussion of the various cases,
and in the end we obtain:
\begin{eqnarray}
&&
-\frac{g^4}{32\pi p}\!\times\!\left\{
\begin{aligned}
&\left[\Lambda_{_{\rm UV}}-\frac{p+|p_0|}{2}\right]
\theta\left(\Lambda_{_{\rm UV}}-\frac{p+|p_0|}{2}\right) &{\scriptstyle[P^2<0]}\\
& \frac{p}{2}
&{\scriptstyle[P^2>0, \frac{p+|p_0|}{2}\le\Lambda_{_{\rm UV}}]}\\
&\Lambda_{_{\rm UV}}-\frac{|p_0|}{2}
&{\scriptstyle [P^2>0, \frac{|p_0|}{2}\le\Lambda_{_{\rm UV}}\le\frac{p+|p_0|}{2}]}\\
& 0
&{\scriptstyle[P^2>0, \Lambda_{_{\rm UV}}\le\frac{|p_0|}{2}]}
\end{aligned}
\right.
\nonumber\\
&&
\end{eqnarray}
instead of eq.~(\ref{eq:1122-cut2}). Note that the ultraviolet
divergence itself (i.e. the terms that diverge when $\Lambda_{_{\rm
    UV}}\to +\infty$ at fixed $P^\mu$) is not affected by this
modification of the regularization procedure.

\bibliographystyle{unsrt}

\begin{thebibliography}{10}

\bibitem{BergeBG1}
{J. Berges, J.P. Blaizot, F. Gelis}, J. Phys. {\bf G 39}, 085115 (2012).

\bibitem{BergeBS1}
{J. Berges, K. Boguslavski, S. Schlichting}, Phys. Rev. {\bf D} {\bf 85},
  076005 (2012).

\bibitem{BergeBSV1}
{J. Berges, K. Boguslavski, S. Schlichting, R. Venugopalan}, arXiv:1303.5650.

\bibitem{BergeBSV2}
{J. Berges, K. Boguslavski, S. Schlichting, R. Venugopalan}, arXiv:1311.3005.

\bibitem{BergeBSV3}
{J. Berges, K. Boguslavski, S. Schlichting, R. Venugopalan}, arXiv:1312.5216.

\bibitem{BergeGSS1}
{J. Berges, D. Gelfand, S. Scheffler, D. Sexty}, Phys. Lett. {\bf B} {\bf 677},
  210 (2009).

\bibitem{BergeS4}
{J. Berges, D. Sexty}, Phys. Rev. Lett. {\bf 108}, 161601 (2012).

\bibitem{BergeSS3}
{J. Berges, S. Schlichting, D. Sexty}, Phys. Rev. {\bf D} {\bf 86}, 074006
  (2012).

\bibitem{FukusG1}
{K. Fukushima, F. Gelis}, Nucl. Phys. {\bf A} {\bf 874}, 108 (2012).

\bibitem{Fukus3}
{K. Fukushima}, arXiv:1307.1046.

\bibitem{KunihMOST1}
{T. Kunihiro, B. Muller, A. Ohnishi, A. Schafer, T.T. Takahashi, {\bf A}
  Yamamoto}, Phys. Rev. {\bf D} {\bf 82}, 114015 (2010).

\bibitem{IidaKMOS1}
{H. Iida, T. Kunihiro, B. Mueller, A. Ohnishi, A. Schaefer, T.T. Takahashi},
  Phys. Rev. {\bf D} {\bf 88}, 094006 (2013).

\bibitem{DusliEGV1}
{K. Dusling, T. Epelbaum, F. Gelis, R. Venugopalan}, Nucl. Phys. {\bf A} {\bf
  850}, 69 (2011).

\bibitem{EpelbG1}
{T. Epelbaum, F. Gelis}, Nucl. Phys. {\bf A} {\bf 872}, 210 (2011).

\bibitem{DusliEGV2}
{K. Dusling, T. Epelbaum, F. Gelis, R. Venugopalan}, Phys. Rev. {\bf D} {\bf
  86}, 085040 (2012).

\bibitem{EpelbG2}
{T. Epelbaum, F. Gelis}, Phys. Rev. {\bf D} {\bf 88}, 085015 (2013).

\bibitem{EpelbG3}
{T. Epelbaum, F. Gelis}, Phys. Rev. Lett. {\bf 111}, 232301 (2013).

\bibitem{KurkeM1}
{A. Kurkela, G.D. Moore}, JHEP {\bf 1112}, 044 (2011).

\bibitem{KurkeM2}
{A. Kurkela, G.D. Moore}, JHEP {\bf 1111}, 120 (2011).

\bibitem{KurkeM3}
{A. Kurkela, G.D. Moore}, Phys. Rev. {\bf D} {\bf 86}, 056008 (2012).

\bibitem{YorkKLM1}
{M.C. Abraao York, A. Kurkela, E. Lu, G.D. Moore}, arXiv:1401.3751.

\bibitem{AttemRS1}
{M. Attems, A. Rebhan, M. Strickland}, Phys. Rev. {\bf D} {\bf 87}, 025010
  (2013).

\bibitem{BlaizGLMV1}
{J.P. Blaizot, F. Gelis, J. Liao, L. McLerran, R. Venugopalan}, Nucl. Phys.
  {\bf A} {\bf 873}, 68 (2012).

\bibitem{BlaizLM2}
{J.P. Blaizot, J. Liao, L.D. McLerran}, arXiv:1305.2119.

\bibitem{HelleJW1}
{M.P. Heller, R.A. Janik, P. Witaszczyk}, Phys. Rev. Lett. {\bf 108}, 201602
  (2012).

\bibitem{CasalHMS1}
{J. Casalderrey-Solana, M.P. Heller, D. Mateos, W. van der Schee}, Phys. Rev.
  Lett. {\bf 111}, 181601 (2013).

\bibitem{Wu1}
{B. Wu}, JHEP {\bf 1304}, 044 (2013).

\bibitem{ReinoS1}
{U. Reinosa, J. Serreau}, Annals Phys. {\bf 325}, 969 (2010).

\bibitem{GautiS1}
{F. Gautier, J. Serreau}, Phys. Rev. {\bf D} {\bf 86}, 125002 (2012).

\bibitem{GirauS1}
{A. Giraud, J. Serreau}, Phys. Rev. Lett. {\bf 104}, 230405 (2010).

\bibitem{FloerW1}
{S. Floerchinger, C. Wetterich}, arXiv:1311.5389.

\bibitem{GasenP1}
{T. Gasenzer, J.M. Pawlowski}, Phys. Lett. {\bf B} {\bf 670}, 135 (2008).

\bibitem{GasenKP1}
{T. Gasenzer, S. Kessler, J.M. Pawlowski}, Eur. Phys. J. {\bf C} {\bf 70}, 423
  (2010).

\bibitem{ArnolMY5}
{P. Arnold, G.D. Moore, L.G. Yaffe}, JHEP {\bf 0301}, 030 (2003).

\bibitem{LuttiW1}
{J.M. Luttinger, J.C. Ward}, Phys. Rev. {\bf 118}, 1417 (1960).

\bibitem{BaymK1}
{G. Baym, L.P. Kadanoff}, Phys. Rev. {\bf 124}, 287 (1961).

\bibitem{DominM1}
{C. de Dominicis, P.C. Martin}, J. Math. Phys. {\bf 5}, 14 (1964).

\bibitem{DominM2}
{C. de Dominicis, P.C. Martin}, J. Math. Phys. {\bf 5}, 31 (1964).

\bibitem{CornwJT1}
{J.M. Cornwall, R. Jackiw, E. Tomboulis}, Phys. Rev. {\bf D} {\bf 10}, 2428
  (1974).

\bibitem{Berge3}
{J. Berges}, AIP Conf. Proc. {\bf 739}, 3 (2005).

\bibitem{BergeBRS1}
{J. Berges, S. Bors{\accent 19 a}nyi, U. Reinosa, J. Serreau}, Annals Phys.
  {\bf 320}, 344 (2005).

\bibitem{ReinoS2}
{U. Reinosa, J. Serreau}, JHEP {\bf 0607}, 028 (2006).

\bibitem{BransG1}
{A. Branschadel, T. Gasenzer}, J. Phys. {\bf B} {\bf 41}, 135302 (2008).

\bibitem{AartsLT1}
{G. Aarts, N. Laurie, A. Tranberg}, Phys. Rev. {\bf D} {\bf 78} 125028 (2008).

\bibitem{HattaN2}
{Y. Hatta, A. Nishiyama}, Nucl. Phys. {\bf A} {\bf 873}, 47 (2012).

\bibitem{HattaN1}
{Y. Hatta, A. Nishiyama}, Phys.Rev. {\bf D} {\bf 86}, 076002 (2012).

\bibitem{IancuLM3}
{E. Iancu, A. Leonidov, L.D. McLerran}, Lectures given at Cargese Summer School
  on QCD Perspectives on Hot and Dense Matter, Cargese, France, 6-18 Aug 2001,
  hep-ph/0202270.

\bibitem{IancuV1}
{E. Iancu, R. Venugopalan}, Quark Gluon Plasma 3, Eds. R.C. Hwa and X.N. Wang,
  World Scientific, hep-ph/0303204.

\bibitem{LappiM1}
{T. Lappi, L.D. McLerran}, Nucl. Phys. {\bf A} {\bf 772}, 200 (2006).

\bibitem{GelisIJV1}
{F. Gelis, E. Iancu, J. Jalilian-Marian, R. Venugopalan}, Ann. Rev. Part. Nucl.
  Sci. {\bf 60}, 463 (2010).

\bibitem{Gelis15}
{F. Gelis}, Int. J. Mod. Phys. {\bf A} {\bf 28}, 1330001 (2013).

\bibitem{PolarS1}
{D. Polarski, A.A. Starobinsky}, Class. Quant. Grav. {\bf 13}, 377 (1996).

\bibitem{Son1}
{D.T. Son}, hep-ph/9601377.

\bibitem{KhlebT1}
{S.Yu. Khlebnikov, I.I. Tkachev}, Phys. Rev. Lett. {\bf 77}, 219 (1996).

\bibitem{GelisLV2}
{F. Gelis, T. Lappi, R. Venugopalan}, Int. J. Mod. Phys. E {\bf 16}, 2595
  (2007).

\bibitem{FukusGM1}
{K. Fukushima, F. Gelis, L. McLerran}, Nucl. Phys. {\bf A} {\bf 786}, 107
  (2007).

\bibitem{Jeon4}
{S. Jeon}, Annals Phys. {\bf 340}, 119 (2014).

\bibitem{BodekMS1}
{D. Bodeker, L.D. McLerran, A. Smilga}, Phys. Rev. {\bf D} {\bf 52}, 4675
  (1995).

\bibitem{ArnolSY1}
{P. Arnold, D.T. Son, L.G. Yaffe}, Phys. Rev. {\bf D} {\bf 55}, 6264 (1997).

\bibitem{AartsS1}
{G. Aarts, J. Smit}, Phys. Lett. {\bf B} {\bf 393}, 395 (1997).

\bibitem{AartsS3}
{G. Aarts, J. Smit}, Nucl. Phys. {\bf B} {\bf 511}, 451 (1998).

\bibitem{AartsNW1}
{G. Aarts, B.J. Nauta, C.G. van Weert}, Phys. Rev. {\bf D} {\bf 61}, 105002
  (2000).

\bibitem{GrigoR1}
{D.Y. Grigoriev, V.A. Rubakov}, Nucl. Phys. {\bf B} {\bf 299}, 67 (1988).

\bibitem{GrigoRS1}
{D.Y. Grigoriev, V.A. Rubakov, M.E. Shaposhnikov}, Nucl. Phys. {\bf B} {\bf
  326}, 737 (1989).

\bibitem{GrigoRS2}
{D.Y. Grigoriev, V.A. Rubakov, M.E. Shaposhnikov}, Phys. Lett. {\bf B} {\bf
  216}, 172 (1989).

\bibitem{AmbjoLS1}
{J. Ambjorn, M. Laursen, M.E. Shaposhnikov}, Phys. Lett. {\bf B} {\bf 197}, 49
  (1987).

\bibitem{AmbjoLS2}
{J. Ambjorn, M. Laursen, M.E. Shaposhnikov}, Nucl. Phys. {\bf B} {\bf 316}, 483
  (1989).

\bibitem{AmbjoAPS2}
{J. Ambjorn, T. Askgaard, H. Porter, M.E. Shaposhnikov}, Phys. Lett. {\bf B}
  {\bf 244}, 479 (1990).

\bibitem{AmbjoAPS1}
{J. Ambjorn, T. Askgaard, H. Porter, M.E. Shaposhnikov}, Nucl. Phys. {\bf B}
  {\bf 353}, 346 (1991).

\bibitem{AmbjoF1}
{J. Ambjorn, K. Farakos}, Phys. Lett. {\bf B} {\bf 294}, 248 (1992).

\bibitem{Schwi1}
{J. Schwinger}, J. Math. Phys. {\bf 2}, 407 (1961).

\bibitem{Keldy1}
{L.V. Keldysh}, Sov. Phys. JETP {\bf 20}, 1018 (1964).

\bibitem{AurenB1}
{P. Aurenche, T. Becherrawy}, Nucl. Phys. {\bf B} {\bf 379}, 259 (1992).

\bibitem{EijckKW1}
{M.A. van Eijck, R. Kobes, Ch.G. van Weert}, Phys. Rev. {\bf D} {\bf 50}, 4097
  (1994).

\bibitem{Gelis3}
{F. Gelis}, Nucl. Phys. {\bf B} {\bf 508}, 483 (1997).

\bibitem{GelisLV3}
{F. Gelis, T. Lappi, R. Venugopalan}, Phys. Rev. {\bf D} {\bf 78}, 054019
  (2008).

\bibitem{GelisLV4}
{F. Gelis, T. Lappi, R. Venugopalan}, Phys. Rev. {\bf D} {\bf 78}, 054020
  (2008).

\bibitem{EpelbGTW1}
{T. Epelbaum, F. Gelis, N. Tanji, B. Wu}, In preparation.

\bibitem{Kubo1}
{R. Kubo}, J. Phys. Soc. Japan {\bf 12}, 570 (1957).

\bibitem{MartiS1}
{P.C. Martin, J. Schwinger}, Phys. Rev. {\bf 115}, 1342 (1959).

\bibitem{Dyson2}
{F.J. Dyson}, Phys. Rev. {\bf 75}, 1736 (1949).

\bibitem{Weinb4}
{S. Weinberg}, Phys. Rev. {\bf 111}, 838 (1960).

\bibitem{HahnZ1}
{Y. Hahn, W. Zimmermann}, Comm. Math. Phys. {\bf 10}, 330 (1968).

\bibitem{Zimme1}
{W. Zimmermann}, Comm. Math. Phys. {\bf 11}, 1 (1968).

\bibitem{CasweK1}
{W.E. Caswell, A.D. Kennedy}, Phys. Rev. {\bf D} {\bf 25}, 392 (1982).

\bibitem{CasweK2}
{W.E. Caswell, A.D. Kennedy}, Phys. Rev. {\bf D} {\bf 28}, 3073 (1983).

\bibitem{Colli1}
{J.C. Collins}, {\sl Renormalization}, Cambridge University Press (1984).

\bibitem{Cutko1}
{R.E. Cutkosky}, J. Math. Phys. {\bf 1}, 429 (1960).

\bibitem{t'HooV1}
{G. t'Hooft, M.J.G. Veltman}, CERN report 73-9.

\bibitem{Hubba1}
{J. Hubbard}, Phys. Rev. Lett. {\bf 3}, 77 (1959).

\bibitem{Strat1}
{R.L. Stratonovich}, Sov. Phys. Dok. {\bf 2} 416 (1957).

\end{thebibliography}

\end{document}